\documentclass[12pt]{report}
\usepackage[utf8]{inputenc}
\usepackage{geometry}
\usepackage{tikz}
\usepackage{titlesec}
\usepackage{natbib}
\usepackage{textcomp}
\newcommand{\degree}{\ensuremath{^\circ}}
\titleformat{\chapter}[block]
  {\normalfont\huge\bfseries}{\thechapter}{1em}{}
\usepackage{booktabs} 
\usepackage{multirow} 
\usepackage{array}    
\usepackage{epigraph}
\usepackage{graphicx}
\usepackage{subcaption}
\usepackage{amsmath}
\usepackage{amsfonts}
\usepackage{hyperref}
\usepackage{fancyhdr}
\usepackage{titlesec}
\usepackage{anyfontsize}
\usepackage{verbatim}
\usepackage{float}
\usepackage{placeins}
\usepackage{booktabs}
\usepackage{array}
\usepackage{caption}
\usepackage{siunitx}

\fancypagestyle{titlepage}{
  \fancyhf{} 
}

\title{Tropopause Detection and Analysis in the Moroccan Region}
\author{Mohammed El Abdioui}
\date{\today}

\makeatletter
\renewcommand{\maketitle}{
  \begin{titlepage}
    \thispagestyle{titlepage}
    
    \vspace*{-2.5cm}
    \begin{center}
      \begin{minipage}{0.9\textwidth}
        \centering
        \includegraphics[height=4cm]{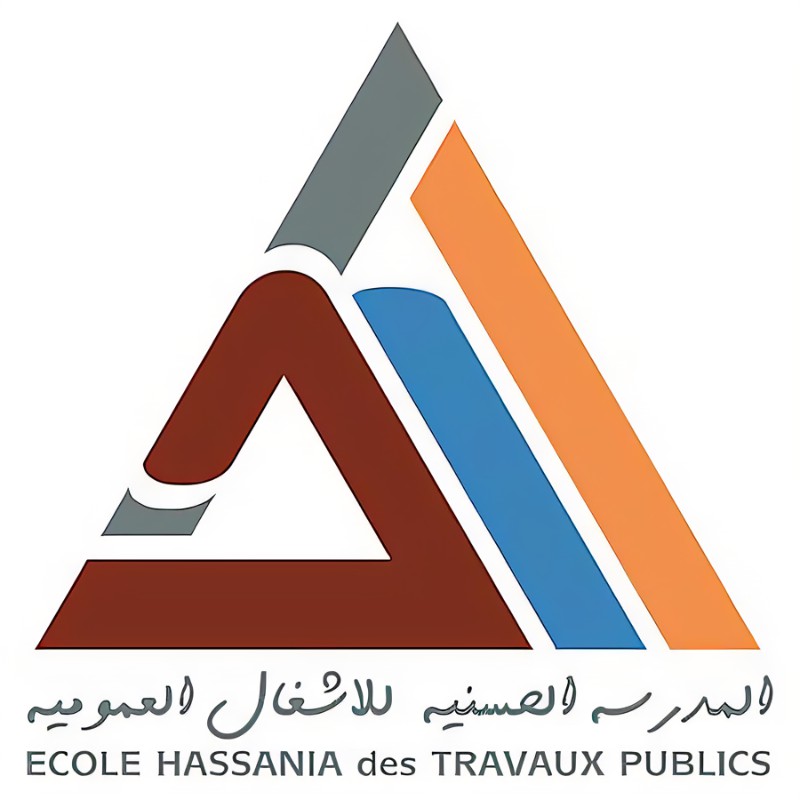} \hfill 
        \includegraphics[height=4cm]{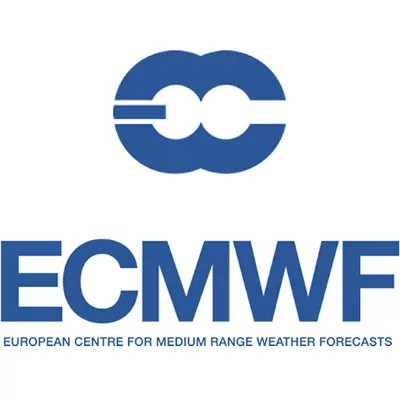} \hfill
        \includegraphics[height=4cm]{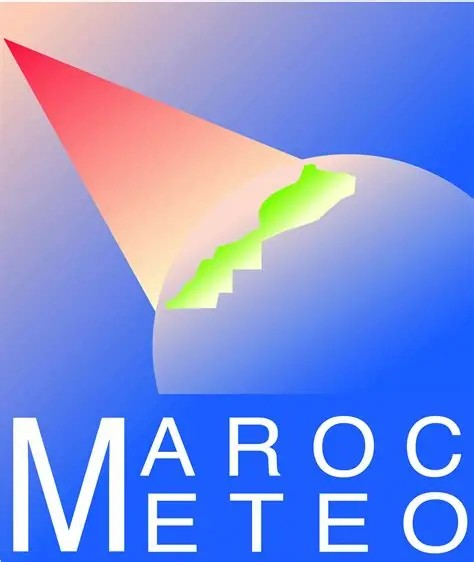}
      \end{minipage}
    \end{center}
    
    \vspace{2cm}
    
    \begin{center}
      {\fontsize{24}{28}\selectfont\textbf{Project Report on Modelisation}}
      
      \vspace{0.5cm}
      
      {\fontsize{20}{24}\selectfont\textbf{\@title}} 
      
      \vspace{0.3cm}
      
      \rule{\textwidth}{1pt}
      
      \vspace{1.5cm}
      
      {\large
      \@author\\
      2nd Year Meteorology Engineering\\
      Hassania School of Public Works (EHTP)
      }
      
      \vspace{1.5cm}
      
      {\large \@date}
      
      \vspace{2cm}
      
      \textit{Supervised by:}\\
      Mr. Noureddine Semane\\ 
      Mr. Khalid El Rhaz\\
      Mrs. Zahra Sahlaoui\\ 
      and Mrs. Siham Sbi\\
    \end{center}
  \end{titlepage}
}
\makeatother

\begin{document}

\maketitle

\newpage
\chapter*{Acknowledgments}

First and foremost, I would like to express my deepest gratitude to my supervisor, \textbf{Mr. Noureddine Semane}, for his invaluable guidance, unwavering support, and dedication throughout the duration of this project. His insightful feedback, patience, and encouragement have been essential in shaping this work. I am also incredibly thankful to \textbf{Mr. Khalid El Rhaz}, \textbf{Mrs. Zahra Sahlaoui}, and \textbf{Mrs. Siham Sbi} for their continuous guidance and support. Each of them brought their unique expertise, which has been crucial in the completion of this project. Their constructive criticism and thought-provoking suggestions allowed me to refine my analysis and improve the overall quality of my work. The collective belief they all showed in my abilities has been a great source of motivation throughout this journey.

I am particularly grateful to my institution, \textbf{Hassania School of Public Works (EHTP)}, for providing the academic environment that encouraged my growth as a researcher. The exposure to advanced meteorological concepts and modeling techniques offered by the faculty has been an invaluable part of my academic journey.

I would like to extend my special thanks to the \textbf{National Directorate of Meteorology (DGM)} and the \textbf{European Centre for Medium-Range Weather Forecasts (ECMWF)}. Their cooperation and the access to high-quality meteorological data made this research possible. The resources and tools made available by these organizations were fundamental in the analysis and modeling stages of this project.

A heartfelt thank you goes to my colleagues and friends for their camaraderie, encouragement, and valuable discussions. Their input and insights were helpful in broadening my perspective and ensuring the accuracy and relevance of my work. I am also grateful to the technical staff at EHTP and DGM for providing the support necessary to access and process the required data.

Finally, I wish to express my profound appreciation to my \textbf{family}, especially my \textbf{parents}, for their continuous love and encouragement. Their belief in me and their unwavering support have been the pillars upon which I have built my success. Without their constant motivation and sacrifices, I would not have been able to reach this stage.

This project is a testament to the collective effort and support of all the people mentioned above, and to those who have contributed in various ways.

\chapter*{Abstract}
This study presents an advanced framework for tropopause detection and analysis using ERA5 reanalysis data, with particular application to extreme meteorological events affecting Morocco and Southern Europe. The research implements and compares multiple detection methodologies, including classical approaches (thermal/WMO and dynamical/1.5 PVU criteria) alongside novel hybrid techniques combining stability and humidity parameters originally developed by ECMWF. 

Through systematic validation against January 2010 monthly means and a detailed case study of the October 2024 DANA (Depresión Aislada en Niveles Altos) event, this work demonstrates the superior performance of hybrid approaches in capturing tropopause dynamics. The development of an automated Python-based comparison tool enabled a quantitative evaluation against Water Vapor 6.2 $\mu$m satellite imagery, revealing that the hybrid method achieved the highest structural similarity (mean SSIM: 0.5327) and the most consistent performance across the studied meteorological conditions.

The findings highlight the critical role of accurate tropopause detection in improving forecasting capabilities for extreme weather events, particularly in topographically complex regions like Southeast Morocco. This research contributes both methodological advancements in tropopause identification and practical insights for operational meteorology, providing a foundation for enhanced early warning systems and improved understanding of atmospheric processes governing high-impact weather phenomena.

\vspace{0.5cm}
\noindent\textbf{Keywords:} Tropopause Detection, ERA5 Reanalysis, DANA Events, Extreme Weather Forecasting, Hybrid Methodology, Satellite Validation, Morocco Meteorology.

\tableofcontents
\listoffigures

\newpage
\chapter{Introduction}

Extreme precipitation events, intensified by climate change and atmospheric variability, present significant challenges for vulnerable regions. These high-intensity, short-duration phenomena can devastate agriculture, infrastructure, and water management systems.  

Southern Morocco, particularly its topographically complex areas, remains highly vulnerable due to limited observations and monitoring networks. In such regions, the lack of reliable data makes it difficult to properly assess and predict extreme weather impacts. At the same time, isolated upper-level low-pressure systems, known as \textbf{DANAs (Depresión Aislada en Niveles Altos)}, play a central role in shaping precipitation extremes over both Europe and North Africa. The October 2024 DANA event, for instance, highlighted the transboundary nature of these phenomena, simultaneously affecting the Iberian Peninsula and Morocco.  

\begin{figure}[H]
    \centering
    \begin{minipage}{0.45\textwidth}
        \centering
        \includegraphics[width=\linewidth]{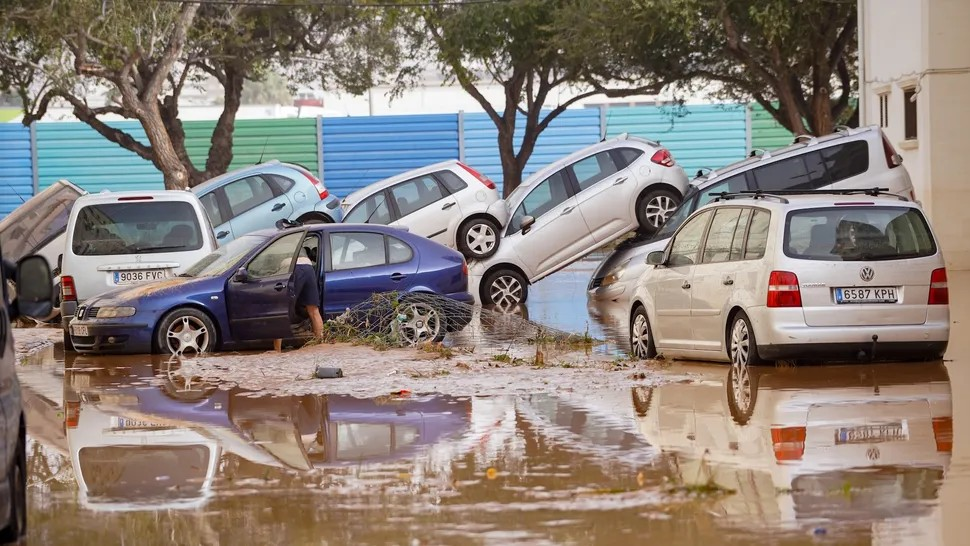}
    \end{minipage}\hfill
    \begin{minipage}{0.45\textwidth}
        \centering
        \includegraphics[width=\linewidth]{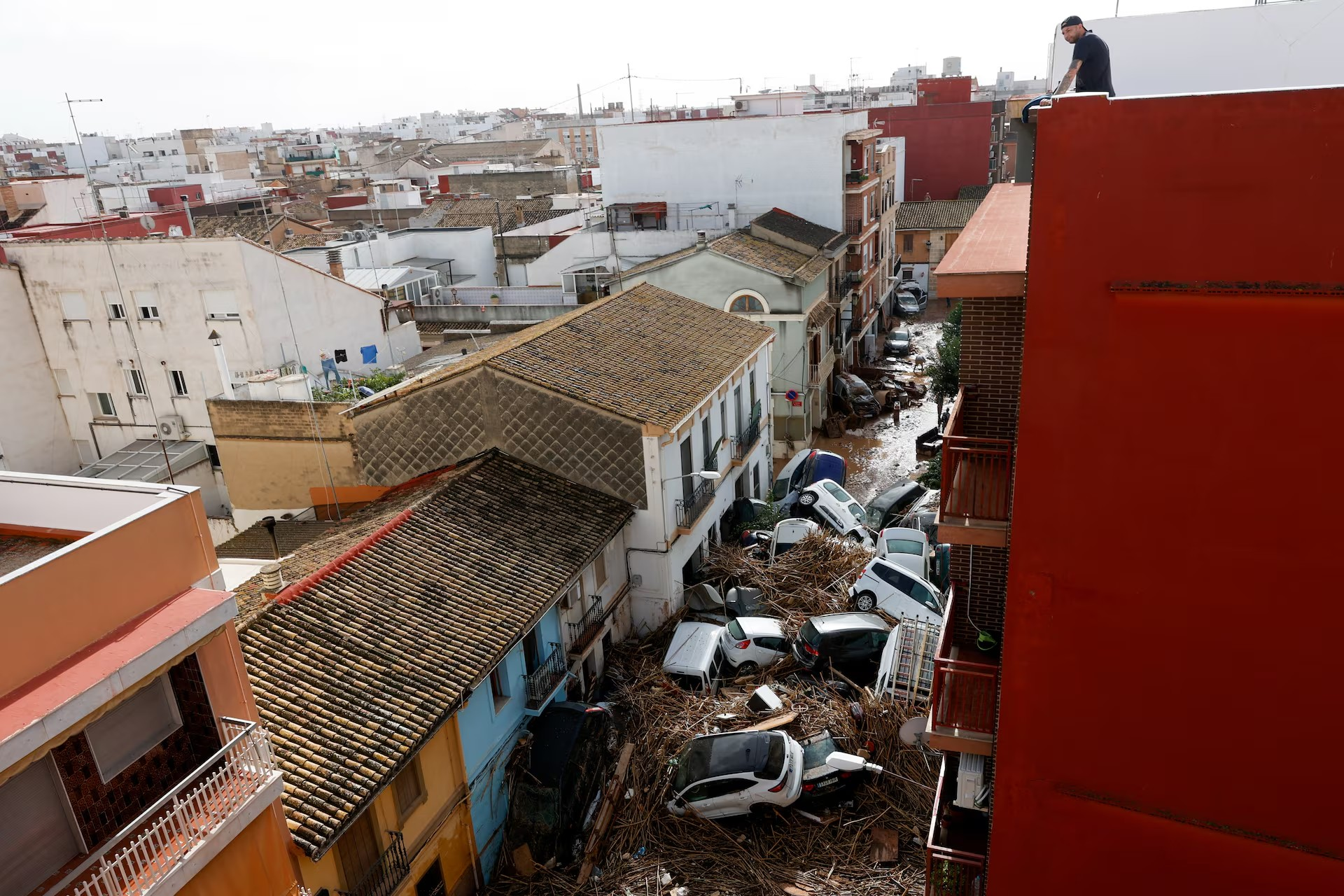}
    \end{minipage}
    \caption{Flood impacts in Valencia, Spain, during the October 2024 DANA event.}
    \label{fig:valencia_floods}
\end{figure}

In this context, numerical weather prediction models such as the \textbf{IFS} global model and the high-resolution limited-area model \textbf{AROME} provide essential tools for understanding extreme weather dynamics, particularly through improved tropopause characterization. This research aims to move beyond traditional potential vorticity-based approaches \cite{potential_vorticity} by implementing innovative detection methodologies that incorporate \textbf{stability}, \textbf{humidity}, and \textbf{hybrid} criteria, using \textbf{ERA5 reanalysis data}.

The central motivation lies in the limitations of conventional definitions—such as the \textbf{thermal (WMO)} \cite{wmo_tropopause} and \textbf{dynamic (2 PVU)} approaches—which often fail to capture tropopause variability in regions of complex topography or strong atmospheric instability. By integrating multi-criteria detection strategies, this work seeks to improve robustness and accuracy in representing tropopause structures, with direct relevance to forecasting extreme weather events such as flash floods in Morocco and DANA-driven precipitation in Europe.  

The main objectives of this project are:  
\begin{itemize}
    \item To develop and apply methods for accurate tropopause detection using ERA5 meteorological fields, including thermal (WMO), dynamical (2 PVU), and hybrid approaches.  
    \item To investigate the relationship between tropopause height and extreme weather phenomena, particularly DANA systems and their impacts on Europe and North Africa.  
    \item To compare the performance of classical and new detection methods (stability, hygropause, and hybrid criteria) in order to evaluate their robustness and limitations.  
    \item To provide recommendations on how improved detection methods could be integrated into forecasting frameworks to enhance weather prediction reliability in regions with limited observations and high variability.  
\end{itemize}

\chapter{Methods and Data}

\section{Classical Tropopause Detection Methods}

\subsection{Thermal (WMO Lapse-Rate Definition)}
The thermal tropopause is defined following the \textbf{WMO lapse-rate criterion} \cite{wmo_tropopause}. It is the lowest level where the temperature lapse rate decreases to $2~\text{K}~\text{km}^{-1}$ or less, provided that the mean lapse rate between this level and all higher levels within $2~\text{km}$ does not exceed $2~\text{K}~\text{km}^{-1}$.

The lapse rate is expressed as:
\begin{equation}
\Gamma = - \frac{\partial T}{\partial z}, \quad z = \frac{\Phi}{g},
\end{equation}
where $T$ is temperature, $\Phi$ is the geopotential, and $g$ is gravity.

The tropopause corresponds to the first level $z_t$ such that:
\begin{equation}
\Gamma(z_t) \leq 2~\text{K}~\text{km}^{-1},
\end{equation}
and for all $z \in [z_t, z_t + 2~\text{km}]$:
\begin{equation}
\overline{\Gamma(z)} \leq 2~\text{K}~\text{km}^{-1}.
\end{equation}

\subsection{Dynamical (2 PVU Criterion)}
The dynamical tropopause is defined as the isosurface where potential vorticity (PV) equals $2~\text{PVU}$ \cite{potential_vorticity}, with:
\[
1~\text{PVU} = 10^{-6}~\text{K}~\text{m}^2~\text{kg}^{-1}~\text{s}^{-1}.
\]

In ERA5, PV is directly available and only rescaled (multiplied by $10^6$) to obtain PVU units. This method is highly sensitive to stratospheric intrusions during cut-off low situations.

\section{New Tropopause Detection Methods}

\subsection{Stability-Based Approach}
The stability-based definition of the tropopause follows the \textbf{ECMWF IFS Documentation} \cite{ifs_documentation}, formulated using the Brunt-Väisälä frequency. The derivation begins with the identity:
\begin{equation}
P_{\theta} \frac{\partial \theta}{\partial p} = P_{T} \frac{\partial T}{\partial p} - \frac{R_d}{c_p},
\end{equation}
together with:
\begin{equation}
P_{\theta} \frac{\partial \theta}{\partial p} = 
P_{\theta} \frac{\partial \theta}{\partial z} \frac{\partial z}{\partial p}
= - \frac{R_d T}{g^2} \frac{g}{\theta} \frac{\partial \theta}{\partial z}
= - \frac{R_d T}{g^2} N_0^2,
\end{equation}
where $N_0$ is the Brunt-Väisälä frequency.

Using a transition value of $N_0^2 = 2.5 \times 10^{-4}~\text{s}^{-2}$, and starting from 70 hPa and moving downward, the tropopause is defined as the first level where:
\begin{equation}
P_{T} \frac{\partial T}{\partial p} + \frac{R_d T}{g^2} N_0^2 \geq \frac{R_d}{c_p}
\end{equation}
is satisfied. The search is constrained between 70 and 500 hPa.

Compared to the WMO definition, this method is both simpler and more robust, particularly in regions of complex vertical layering.

\subsection{Humidity / Hygropause Criterion}
The hygropause is defined by humidity gradients in the 70-500 hPa layer, following ECMWF's approach to stratospheric humidity analysis \cite{ecmwf_stratosphere}:
\begin{itemize}
    \item Specific humidity at the level exceeds $3~\text{mg}~\text{kg}^{-1}$.
    \item Specific humidity two levels below exceeds $5~\text{mg}~\text{kg}^{-1}$.
\end{itemize}
This method captures the sharp moisture decrease marking the transition between the troposphere and the stratosphere.

\subsection{Hybrid Approach (Stability + Humidity)}
The hybrid method combines stability and hygropause criteria, addressing limitations of each when applied individually \cite{hybrid_method}. By requiring both static stability and humidity thresholds, this approach reduces false detections and provides a more consistent representation of the tropopause. Originally developed at ECMWF, this method has been adapted and applied to ERA5 data.

\section{Data and Tools}
\subsection{ERA5 Reanalysis}
The analysis utilizes \textbf{ERA5 reanalysis} data from the European Centre for Medium-Range Weather Forecasts (ECMWF) \cite{era5_documentation}. This dataset has the following characteristics:
\begin{itemize}
    \item \textbf{Spatial resolution}: $0.25\degree \times 0.25\degree$ (approximately 31 km)
    \item \textbf{Temporal resolution}: 3-hourly data
    \item \textbf{Vertical levels}: 37 pressure levels: 1 hPa, 2 hPa, 3 hPa, 5 hPa, 7 hPa, 10 hPa, 20 hPa, 30 hPa, 50 hPa, 70 hPa, 100 hPa, 125 hPa, 150 hPa, 175 hPa, 200 hPa, 225 hPa, 250 hPa, 300 hPa, 350 hPa, 400 hPa, 450 hPa, 500 hPa, 550 hPa, 600 hPa, 650 hPa, 700 hPa, 750 hPa, 775 hPa, 800 hPa, 825 hPa, 850 hPa, 875 hPa, 900 hPa, 925 hPa, 950 hPa, 975 hPa, 1000 hPa
    \item \textbf{Variables}: Temperature, geopotential, specific humidity, potential vorticity, relative humidity
    \item \textbf{Region}: $20\degree$N to $40\degree$N, $20\degree$W to $10\degree$E (covering Morocco and Southern Europe)
\end{itemize}

\subsection{Model Context}
While the primary analysis uses ERA5 reanalysis data, the methodological framework is designed to be compatible with:
\begin{itemize}
    \item \textbf{IFS global model}: ECMWF's Integrated Forecasting System \cite{ifs_documentation}
    \item \textbf{AROME regional model}: Application of Research to Operations at Mesoscale
\end{itemize}
These models provide context for operational forecasting applications.

\subsection{Python-Based Automated Comparison Framework}
A sophisticated Python-based framework was developed to quantitatively evaluate the different tropopause detection methods:
\begin{itemize}
    \item \textbf{Image Processing Pipeline}: Automated timestamp extraction, grayscale conversion, normalization, and resizing of satellite imagery
    \item \textbf{Similarity Metrics}: Computation of quantitative metrics including Structural Similarity Index (SSIM) \cite{ssim_paper}, Mean Squared Error (MSE) \cite{mse_textbook}, and Normalized Cross-Correlation (NCC) \cite{ncc_paper}
    \item \textbf{Visual Analysis}: Generation of composite figures displaying satellite imagery, method outputs, and difference maps
    \item \textbf{Statistical Evaluation}: Comprehensive statistical analysis and method ranking
\end{itemize}

\subsection{Reference Satellite Data}
Water vapor imagery (6.2 $\mu$m channel) serves as a validation reference for tropopause features, with appropriate limb correction applied where necessary \cite{limb_correction}. This satellite data provides an independent observational benchmark for evaluating the performance of different detection methods.
\chapter{Results and Discussion}

\section{Monthly Mean Tropopause Height}

\subsection{Classical vs. New Methods}
A preliminary validation was performed using the \textbf{monthly mean of January 2010} to compare classical and new tropopause detection methods. This period provides a stable baseline for systematic comparison, reducing transient variability and highlighting methodological differences.

All detection methods were implemented on \textbf{ERA5 pressure levels}, with results presented as latitude-pressure cross sections using a logarithmic pressure axis (1000-10 hPa). The WMO thermal method required computation of lapse rates with respect to altitude but was plotted back in pressure coordinates for consistency.

\begin{figure}[H]
    \centering
    \includegraphics[width=0.8\linewidth]{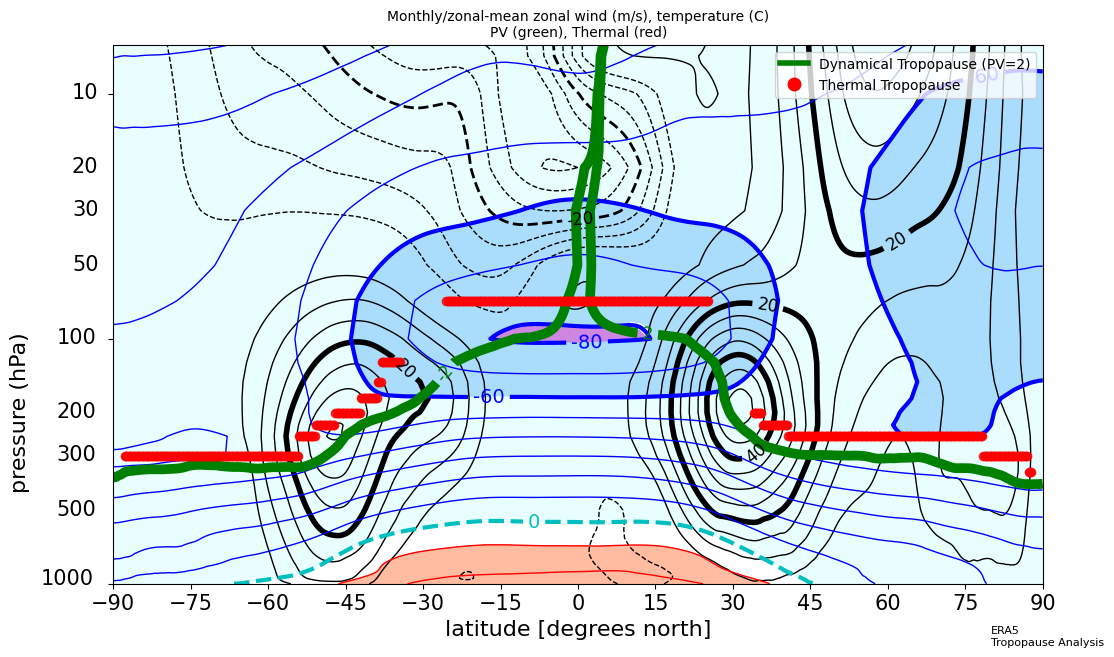}
    \caption{Monthly mean tropopause height (January 2010) using the dynamical (2 PVU) method. The thermal (WMO) tropopause is included as reference.}
    \label{fig:dynamical_method}
\end{figure}

\begin{figure}[H]
    \centering
    \includegraphics[width=0.8\linewidth]{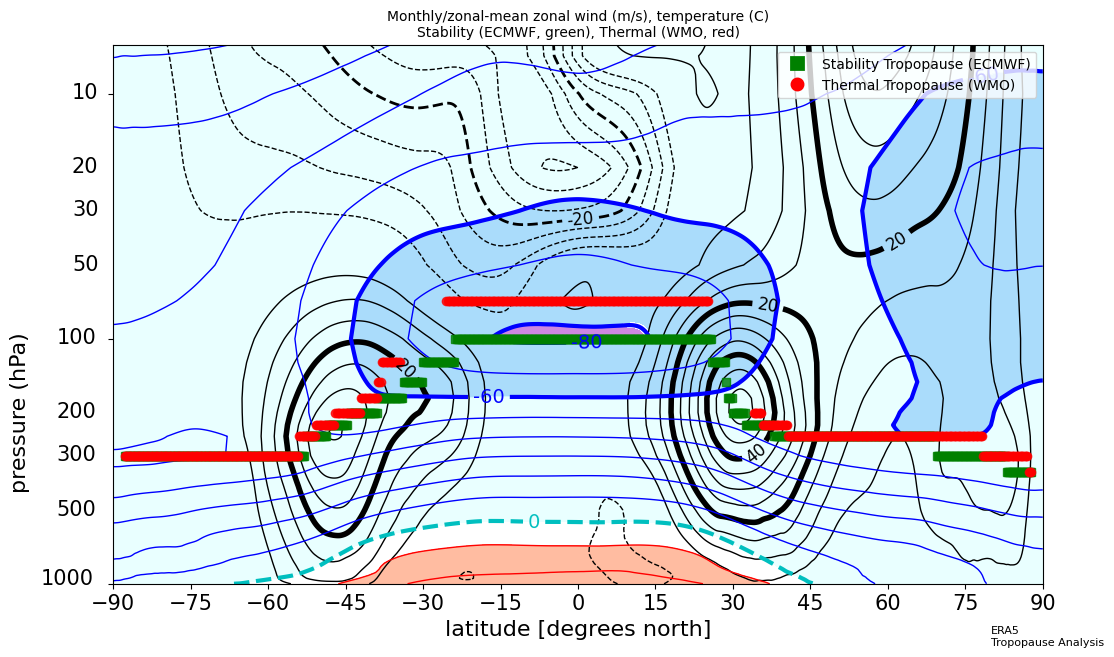}
    \caption{Monthly mean tropopause height (January 2010) using the stability method. The thermal (WMO) tropopause is included as reference.}
    \label{fig:stability_method}
\end{figure}

\begin{figure}[H]
    \centering
    \includegraphics[width=0.8\linewidth]{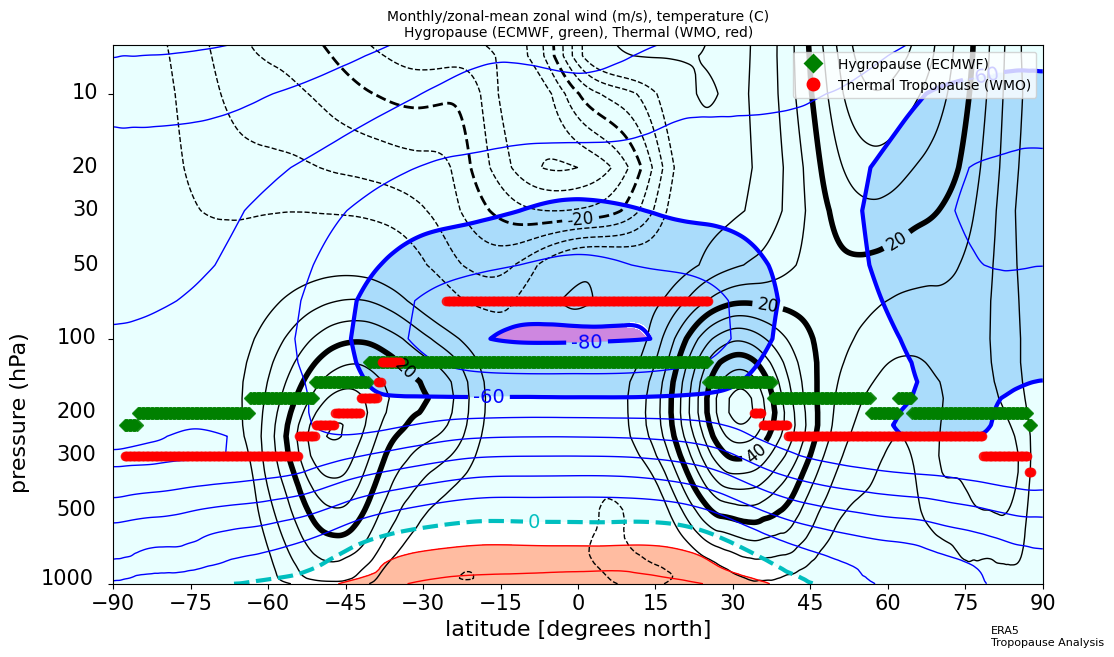}
    \caption{Monthly mean tropopause height (January 2010) using the hygropause method. The thermal (WMO) tropopause is included as reference.}
    \label{fig:hygropause_method}
\end{figure}

\begin{figure}[H]
    \centering
    \includegraphics[width=0.8\linewidth]{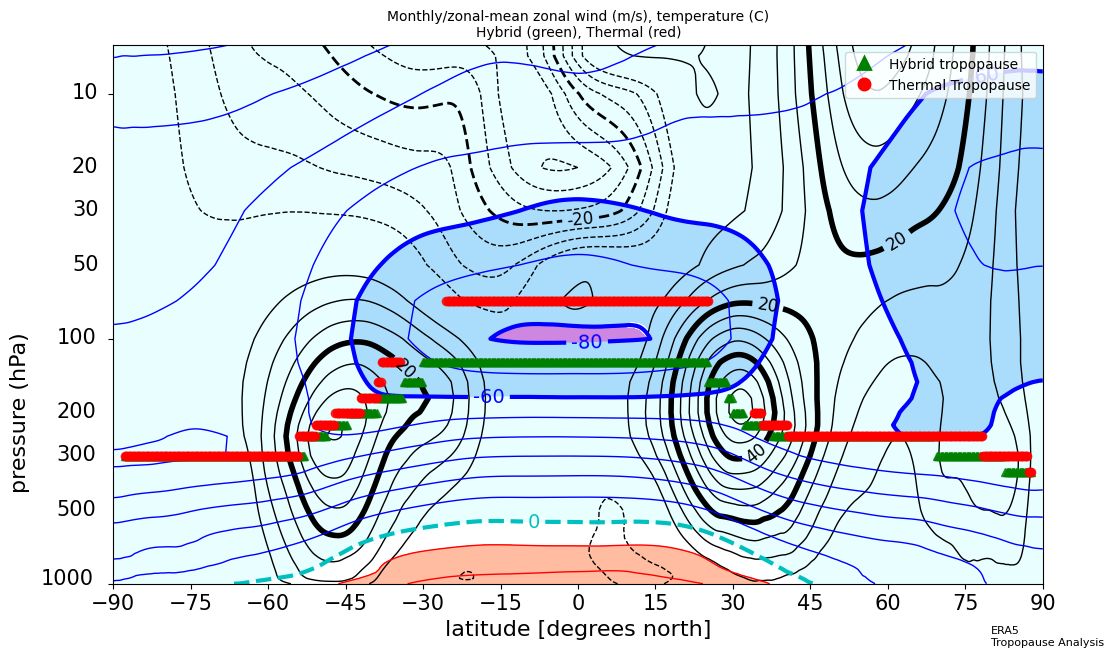}
    \caption{Monthly mean tropopause height (January 2010) using the hybrid method. The thermal (WMO) tropopause is included as reference.}
    \label{fig:hybrid_method}
\end{figure}

\subsection{Systematic Differences}
The monthly mean comparison revealed systematic differences between methods:
\begin{itemize}
    \item The \textbf{thermal (WMO) method} provided a consistent baseline but showed discontinuities in regions of complex vertical structure
    \item The \textbf{dynamical (2 PVU) method} captured stratospheric intrusions effectively but was less sensitive to thermodynamic transitions
    \item The \textbf{stability method} offered improved robustness in regions with complex vertical layering
    \item The \textbf{hygropause method} effectively identified moisture gradients but occasionally failed in dry atmospheric conditions
    \item The \textbf{hybrid method} demonstrated the most consistent performance, combining the strengths of stability and humidity criteria
\end{itemize}

\section{Case Study: DANA (October 2024)}

\subsection{Synoptic Evolution}
The analysis focused on a significant \textbf{DANA (Depresión Aislada en Niveles Altos)} event over the Iberian Peninsula from 27 to 30 October 2024. This event was characterized by the formation of a cut-off low in the upper troposphere, leading to severe rainfall and flooding across Spain, with impacts extending toward southern Morocco.

The large-scale atmospheric setup featured a pronounced upper-level trough with associated stratospheric intrusions, creating favorable conditions for deep convection and extreme precipitation. Water vapor imagery (6.2 $\mu$m) clearly revealed the dry intrusion characteristic of DANA events, providing an excellent validation benchmark.

\subsection{Geographical Maps of Tropopause Height}
Geographical maps of tropopause height were produced for each detection method at three critical stages of the DANA lifecycle: genesis (27 October 00:00 UTC), mature phase (28 October 21:00 UTC), and beginning of dissipation (30 October 21:00 UTC). Results are represented in terms of geopotential height ($z = \Phi/g$) in decameters (dam).

\subsubsection{Dynamical Method (1.5 PVU)}

\begin{figure}[H]
    \centering
    \includegraphics[width=0.6\linewidth]{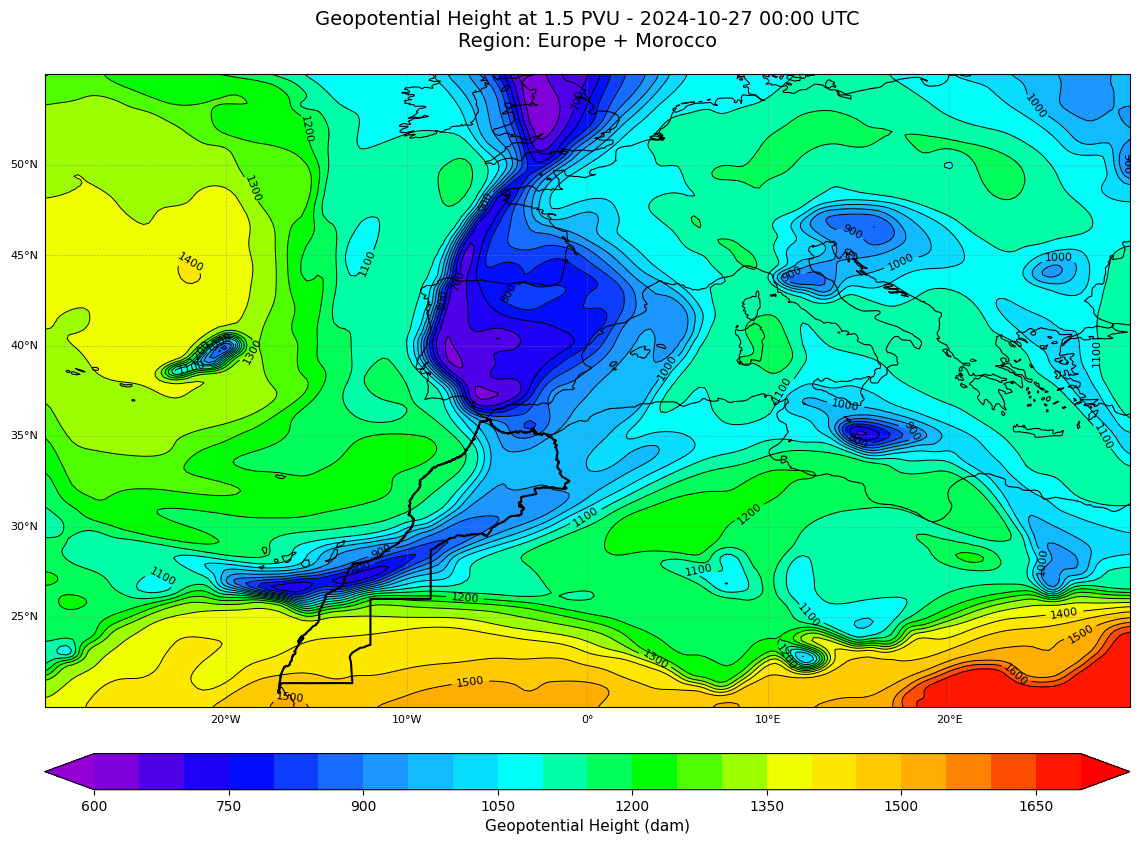}
    \label{fig:dynamical_15_genesis}
\end{figure}

\begin{figure}[H]
    \centering
    \includegraphics[width=0.6\linewidth]{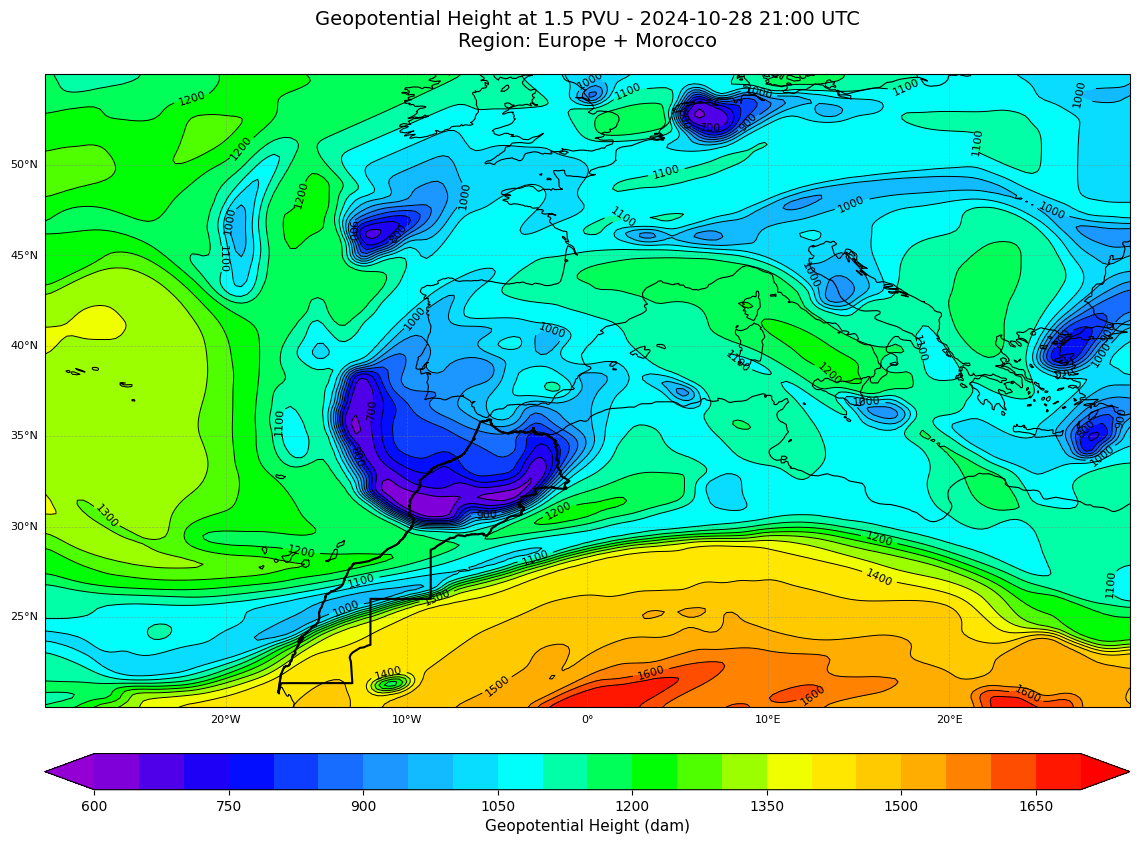}
    \label{fig:dynamical_15_mature}
\end{figure}

\newpage

\begin{figure}[H]
    \centering
    \includegraphics[width=0.6\linewidth]{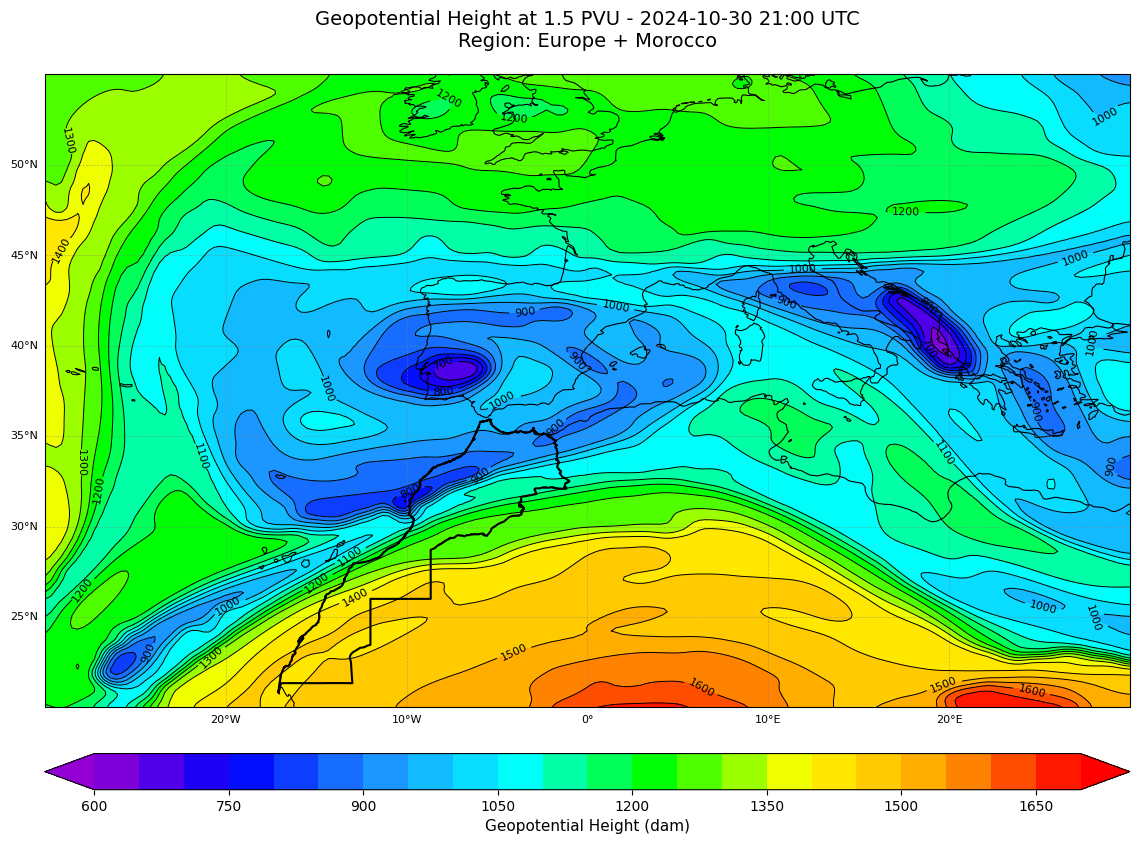}
    \label{fig:dynamical_15_dissipation}
\end{figure}
\vspace{1cm}
\subsubsection{Dynamical Method (2 PVU)}

\begin{figure}[H]
    \centering
    \includegraphics[width=0.6\linewidth]{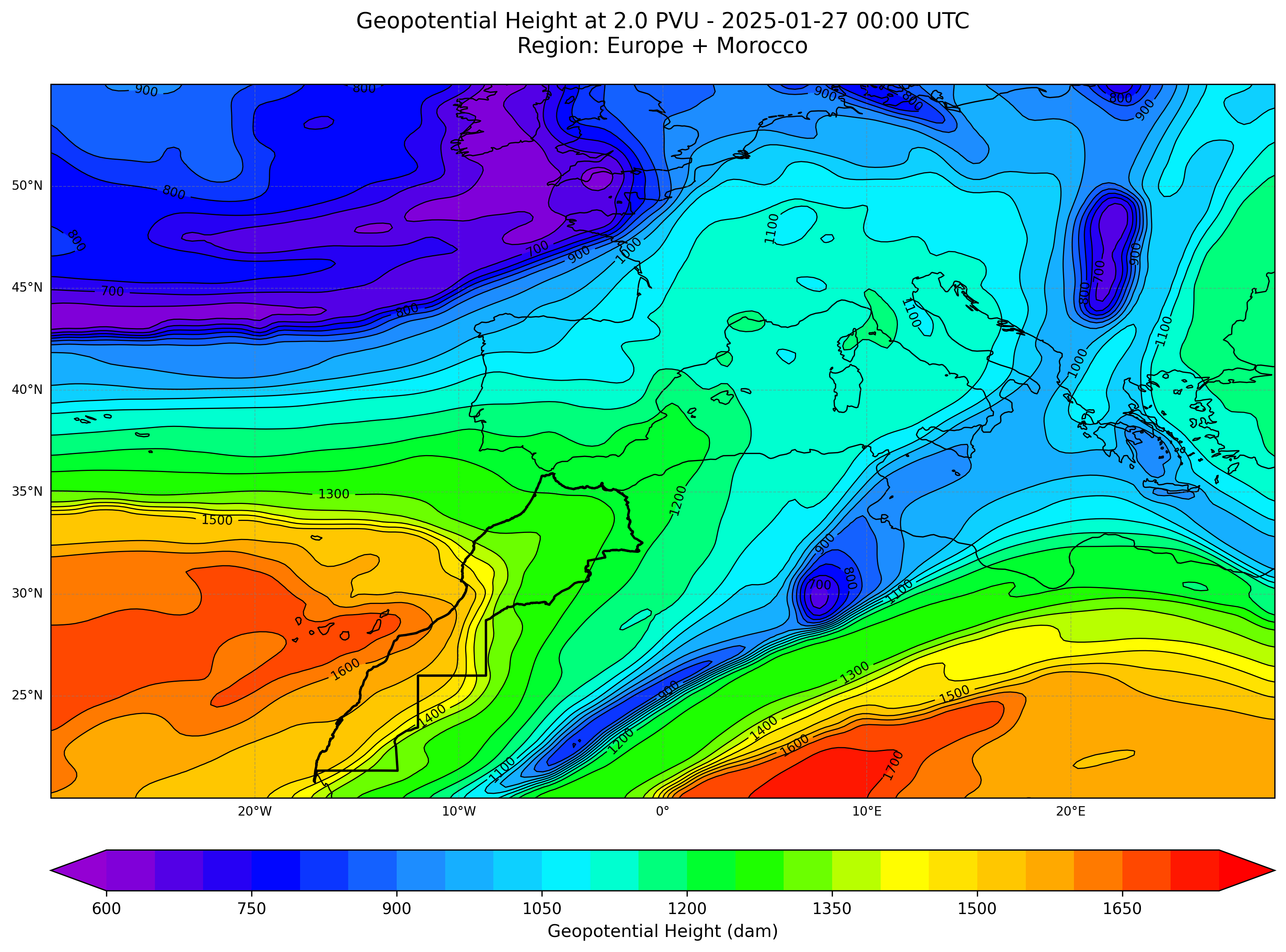}
    \label{fig:dynamical_2_genesis}
\end{figure}

\begin{figure}[H]
    \centering
    \includegraphics[width=0.6\linewidth]{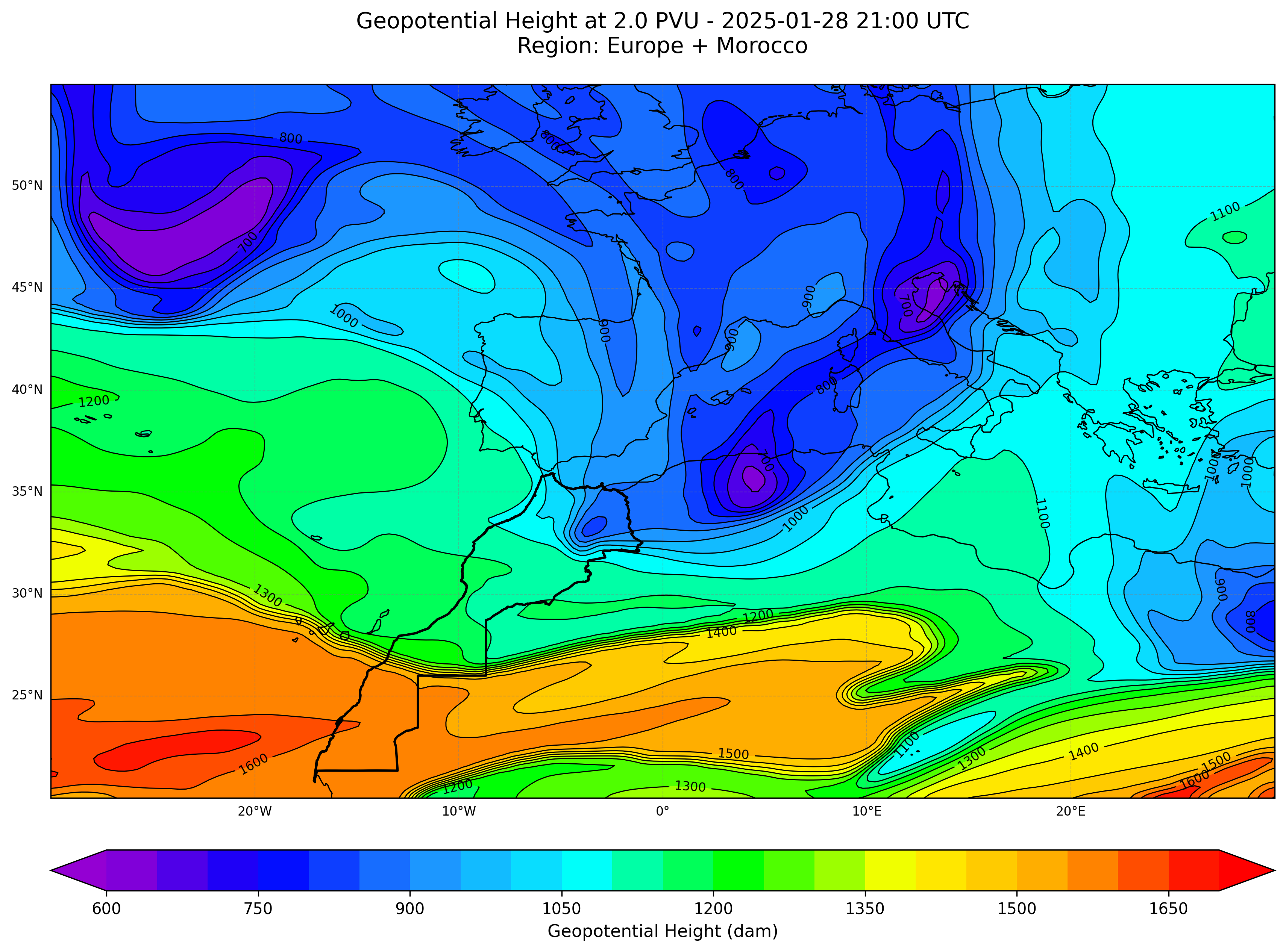}
    \label{fig:dynamical_2_mature}
\end{figure}

\begin{figure}[H]
    \centering
    \includegraphics[width=0.6\linewidth]{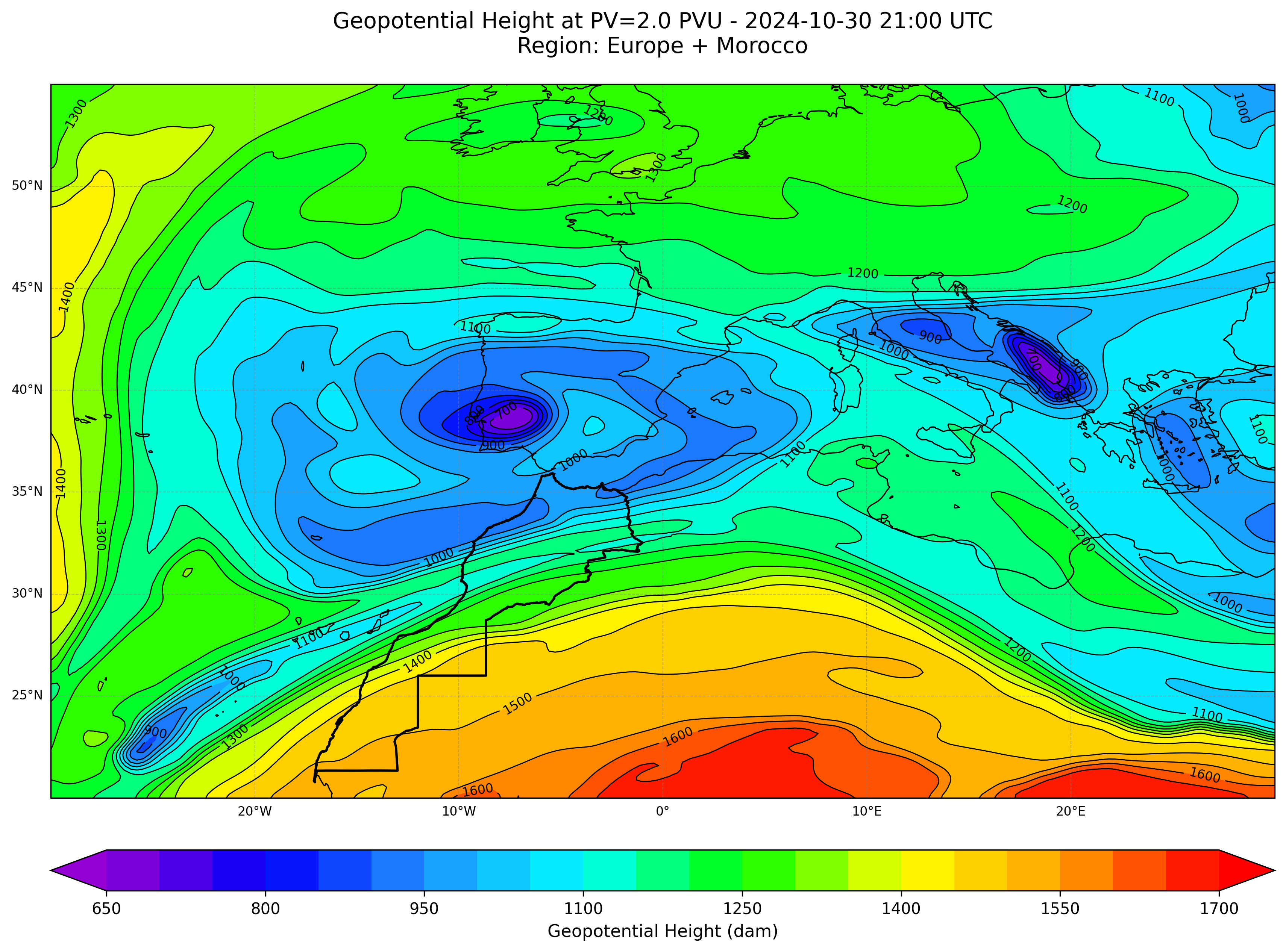}
    \label{fig:dynamical_2_dissipation}
\end{figure}

\subsubsection{Stability Method}

\begin{figure}[H]
    \centering
    \includegraphics[width=0.6\linewidth]{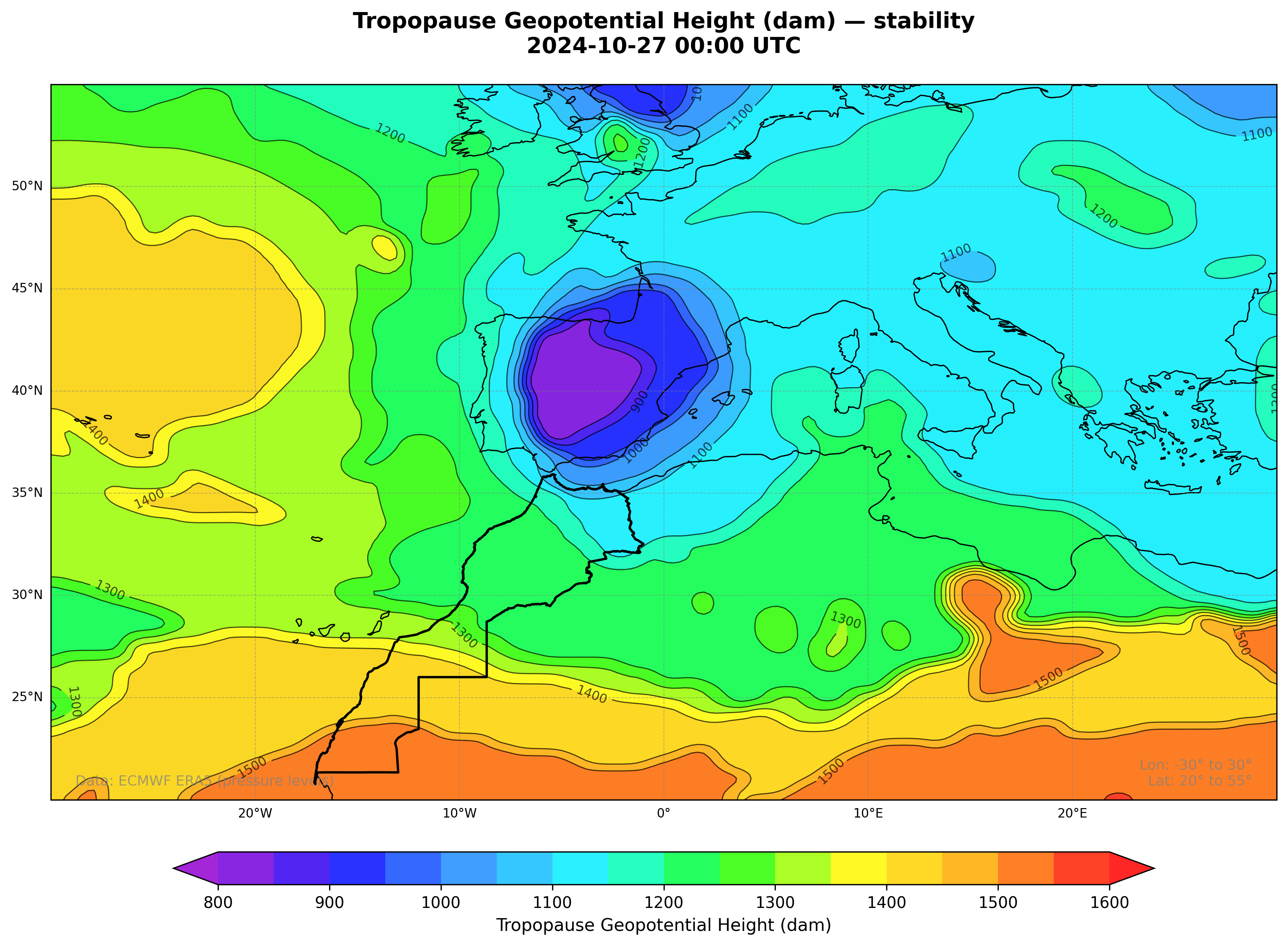}
    \label{fig:stability_genesis}
\end{figure}

\begin{figure}[H]
    \centering
    \includegraphics[width=0.6\linewidth]{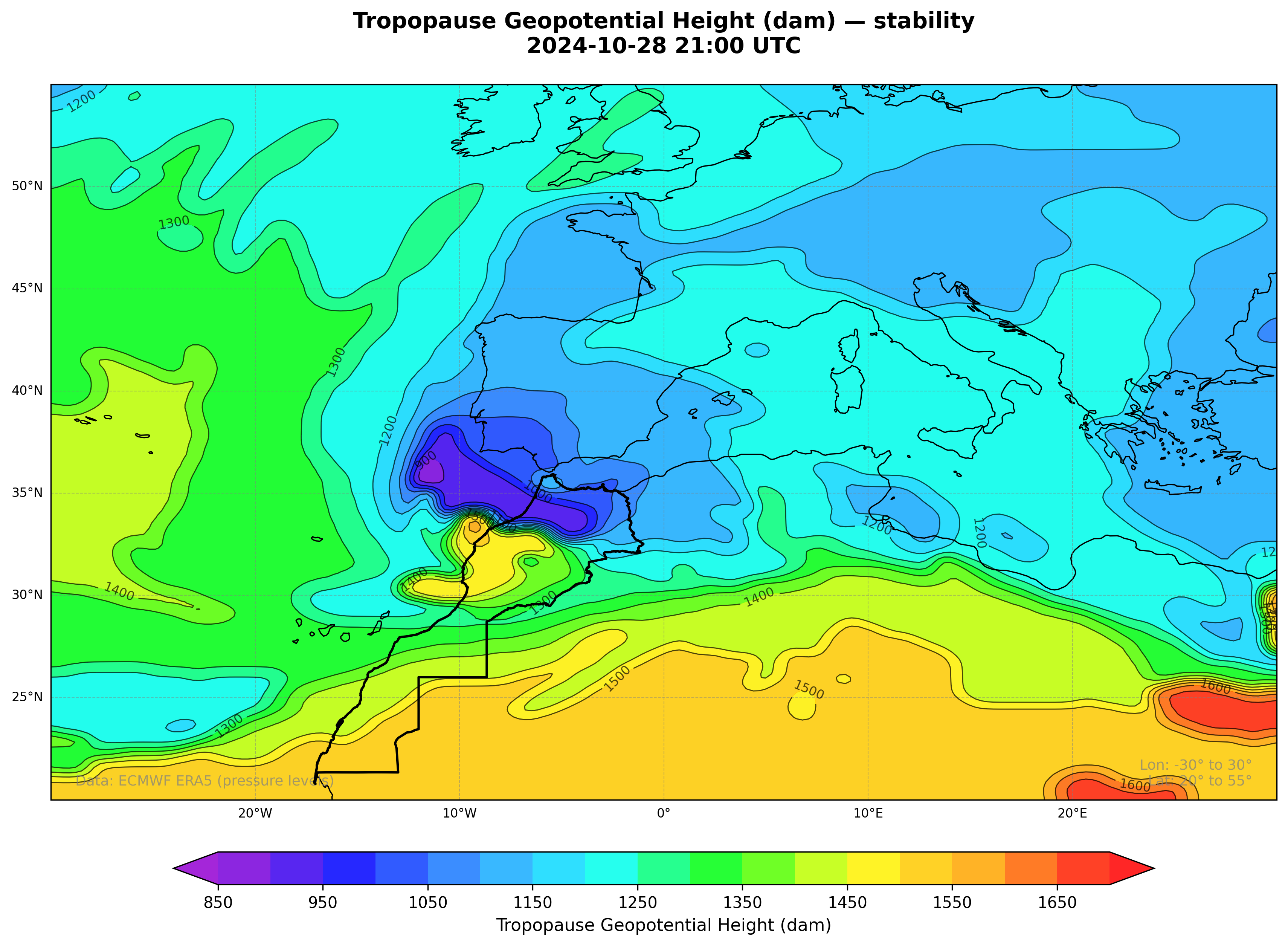}
    \label{fig:stability_mature}
\end{figure}

\newpage

\begin{figure}[H]
    \centering
    \includegraphics[width=0.6\linewidth]{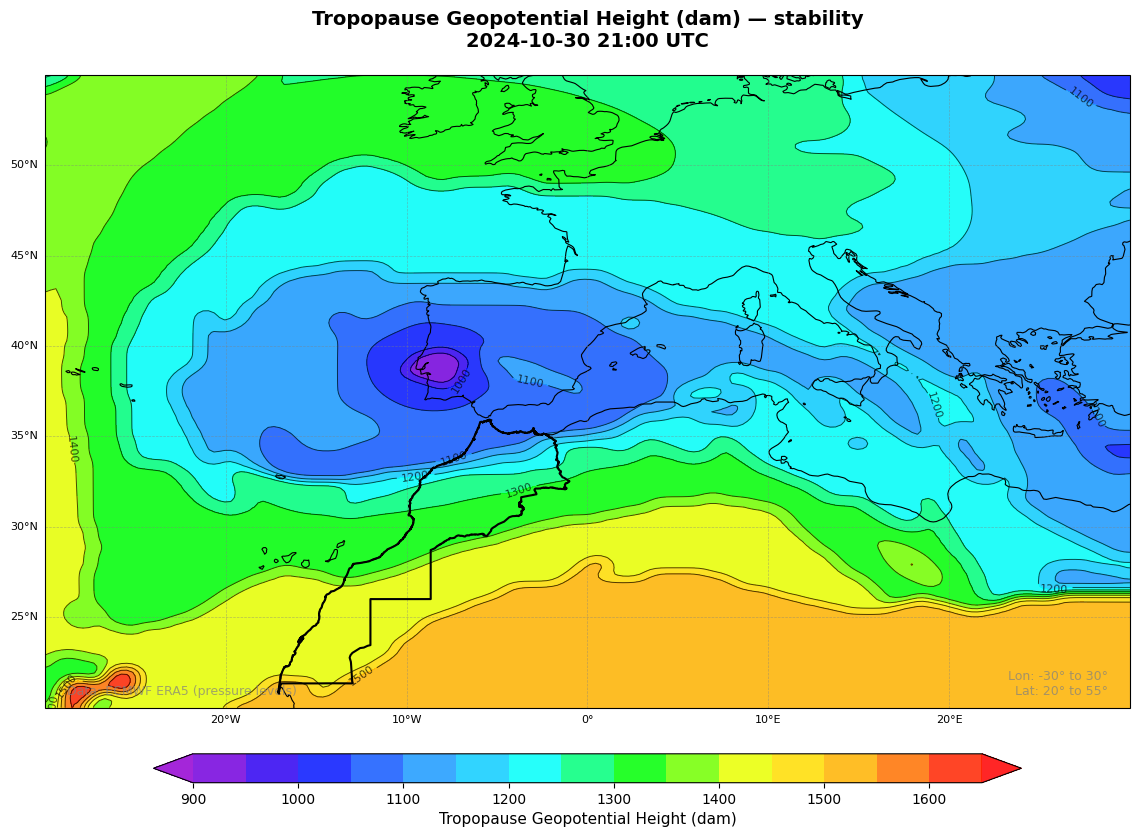}
    \label{fig:stability_dissipation}
\end{figure}
\vspace{1cm}
\subsubsection{Hygropause Method}

\begin{figure}[H]
    \centering
    \includegraphics[width=0.6\linewidth]{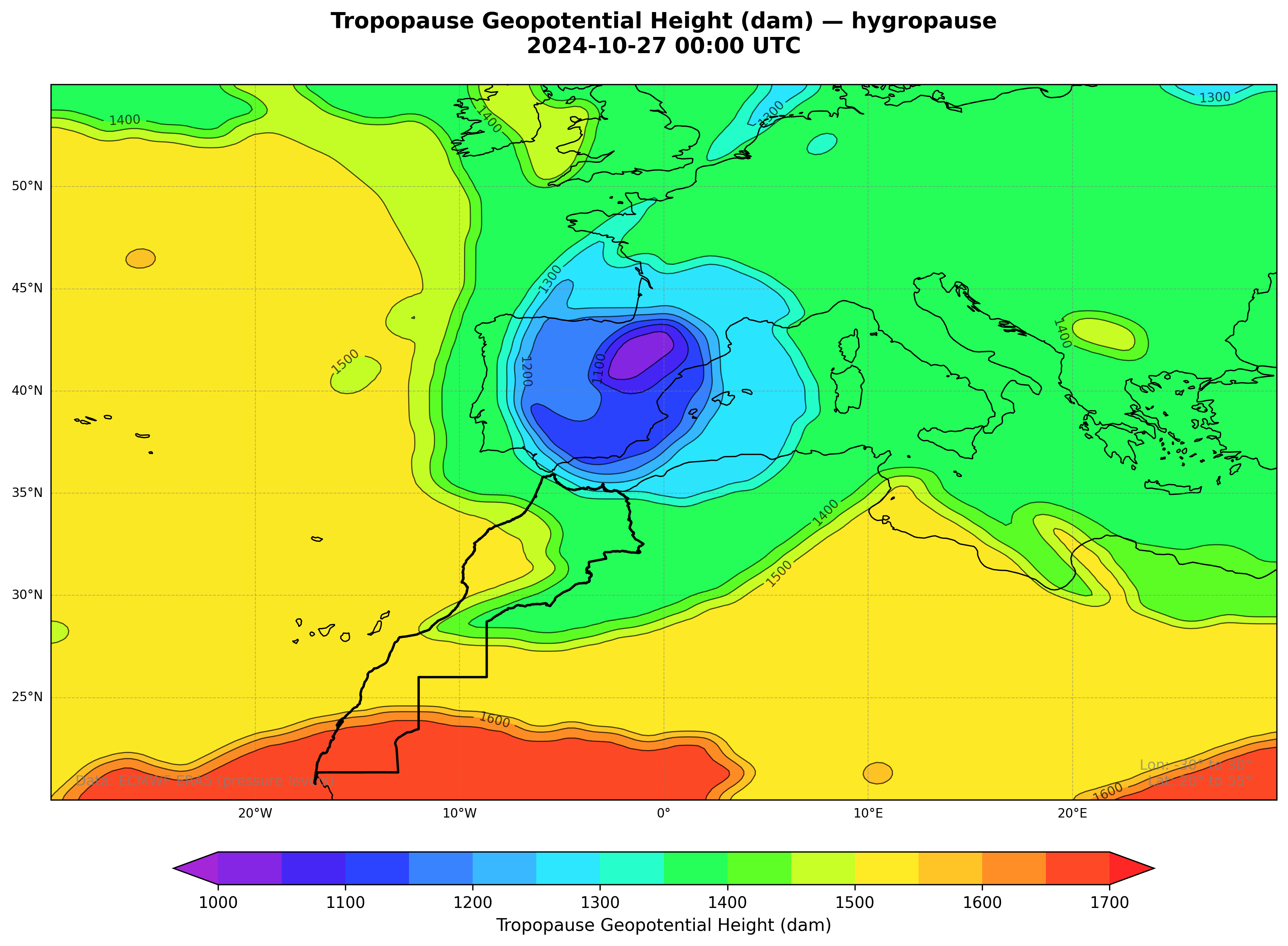}
    \label{fig:hygropause_genesis}
\end{figure}

\begin{figure}[H]
    \centering
    \includegraphics[width=0.6\linewidth]{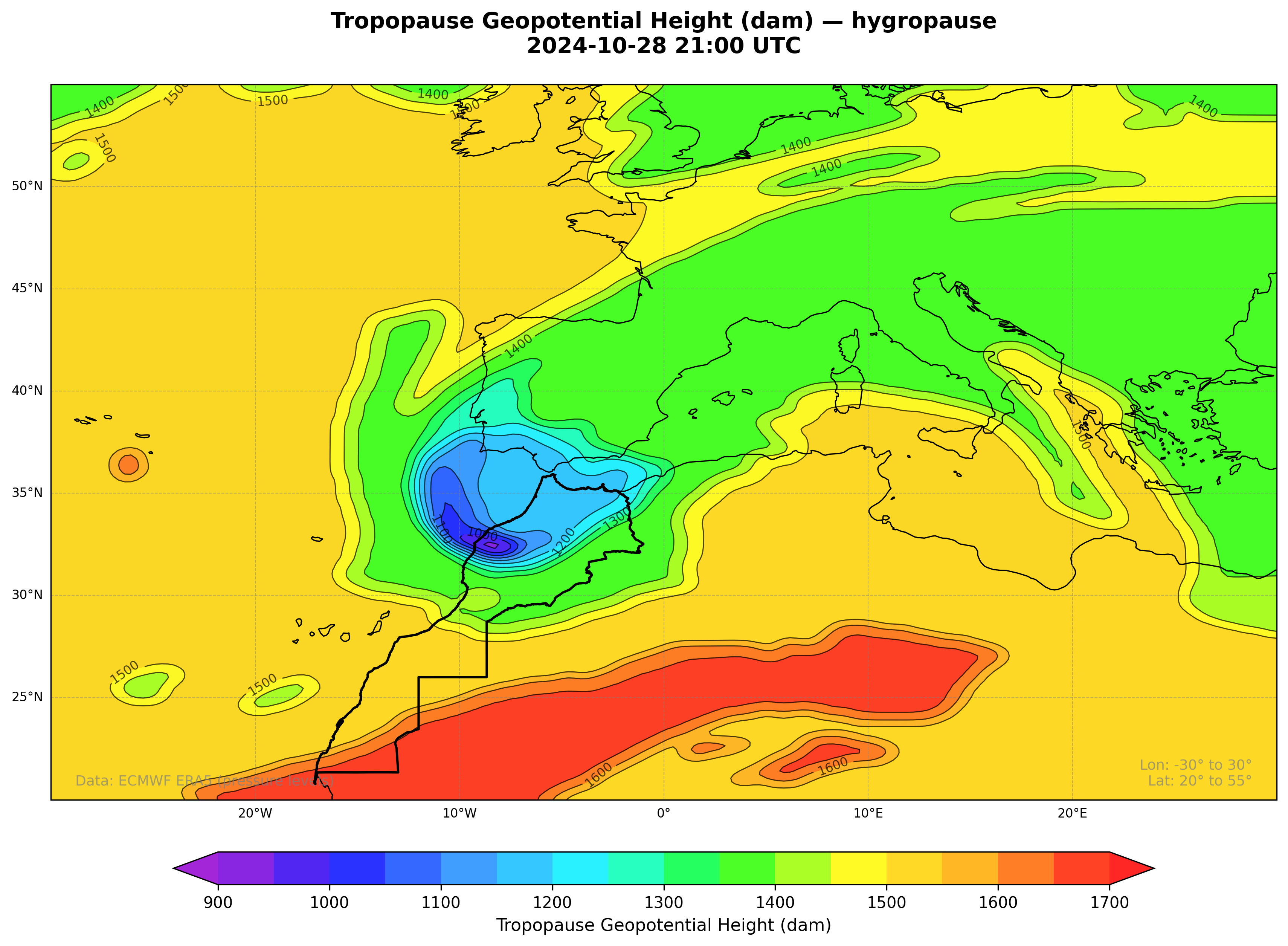}
    \label{fig:hygropause_mature}
\end{figure}

\begin{figure}[H]
    \centering
    \includegraphics[width=0.6\linewidth]{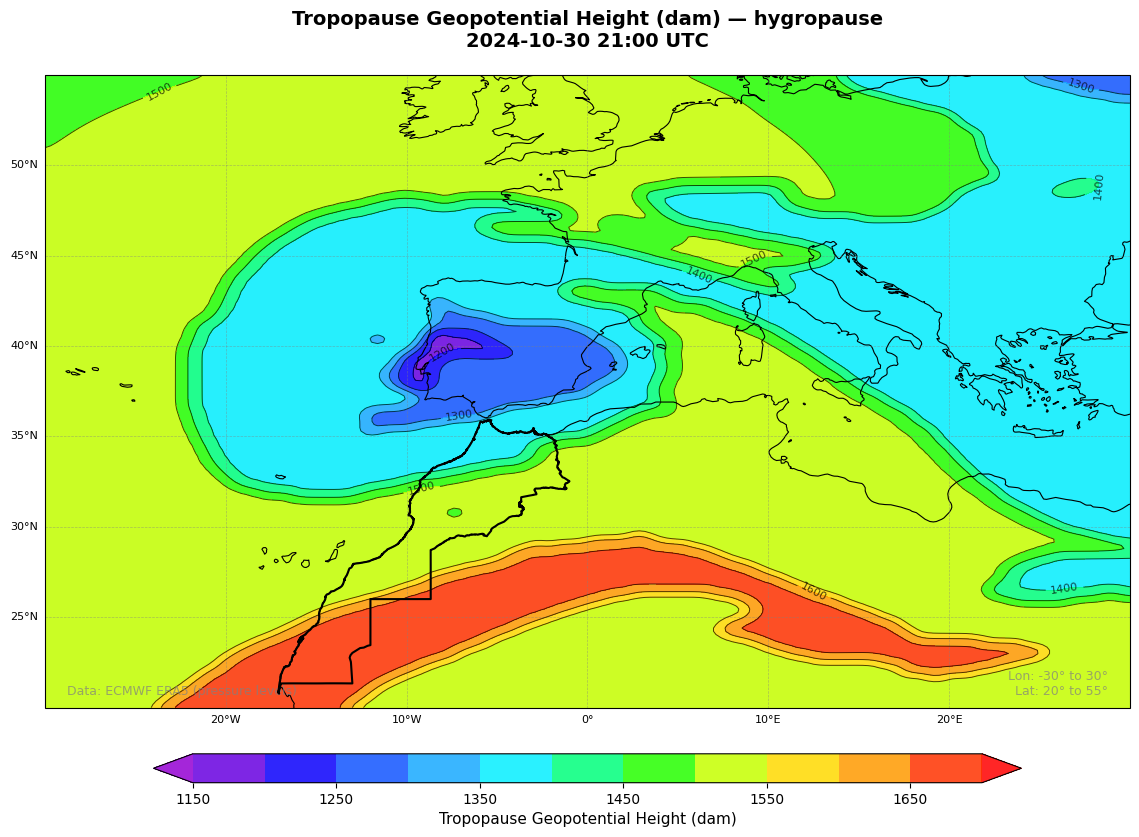}
    \label{fig:hygropause_dissipation}
\end{figure}

\subsubsection{Hybrid Method}

\begin{figure}[H]
    \centering
    \includegraphics[width=0.6\linewidth]{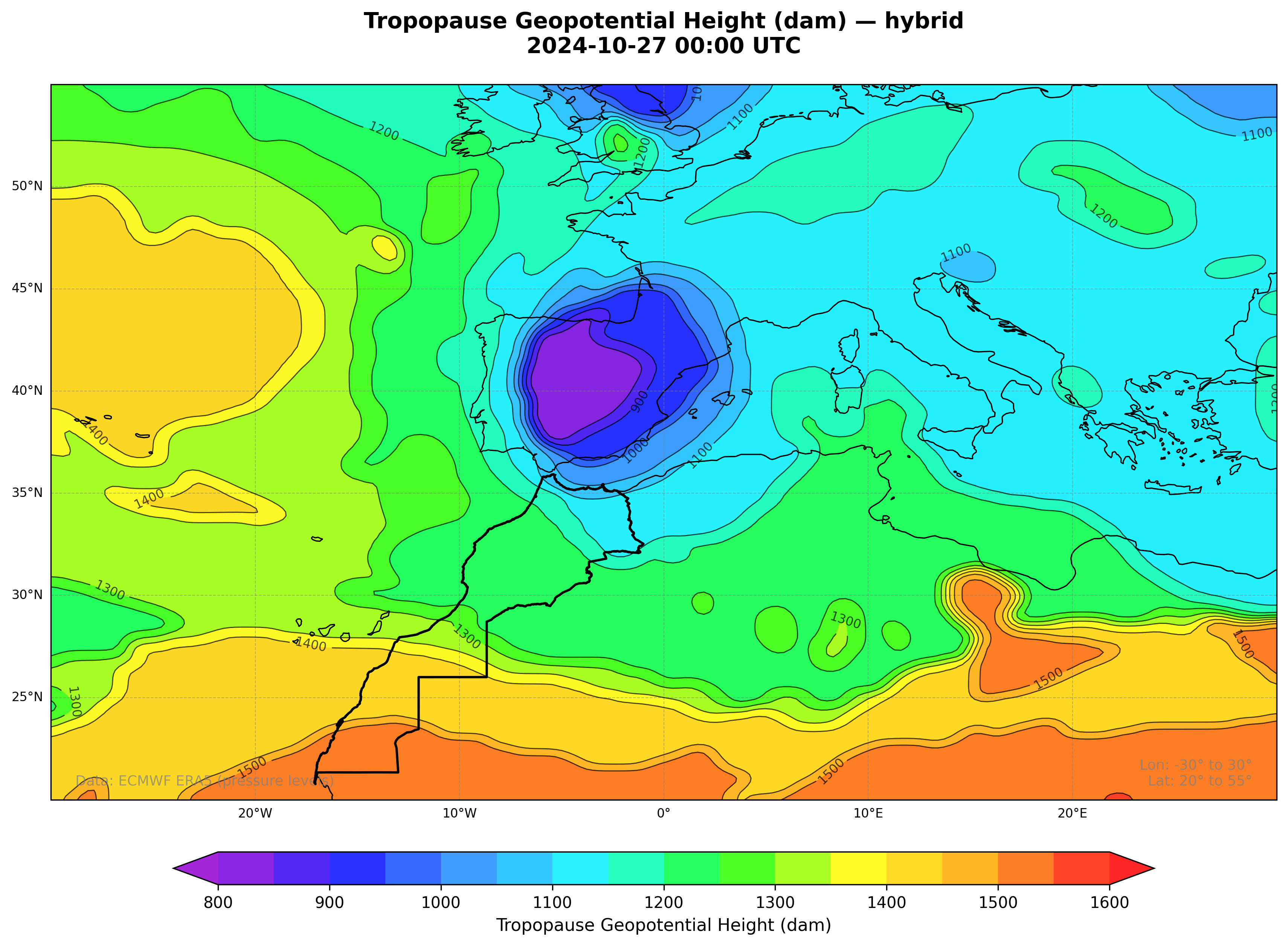}
    \label{fig:hybrid_genesis}
\end{figure}

\begin{figure}[H]
    \centering
    \includegraphics[width=0.6\linewidth]{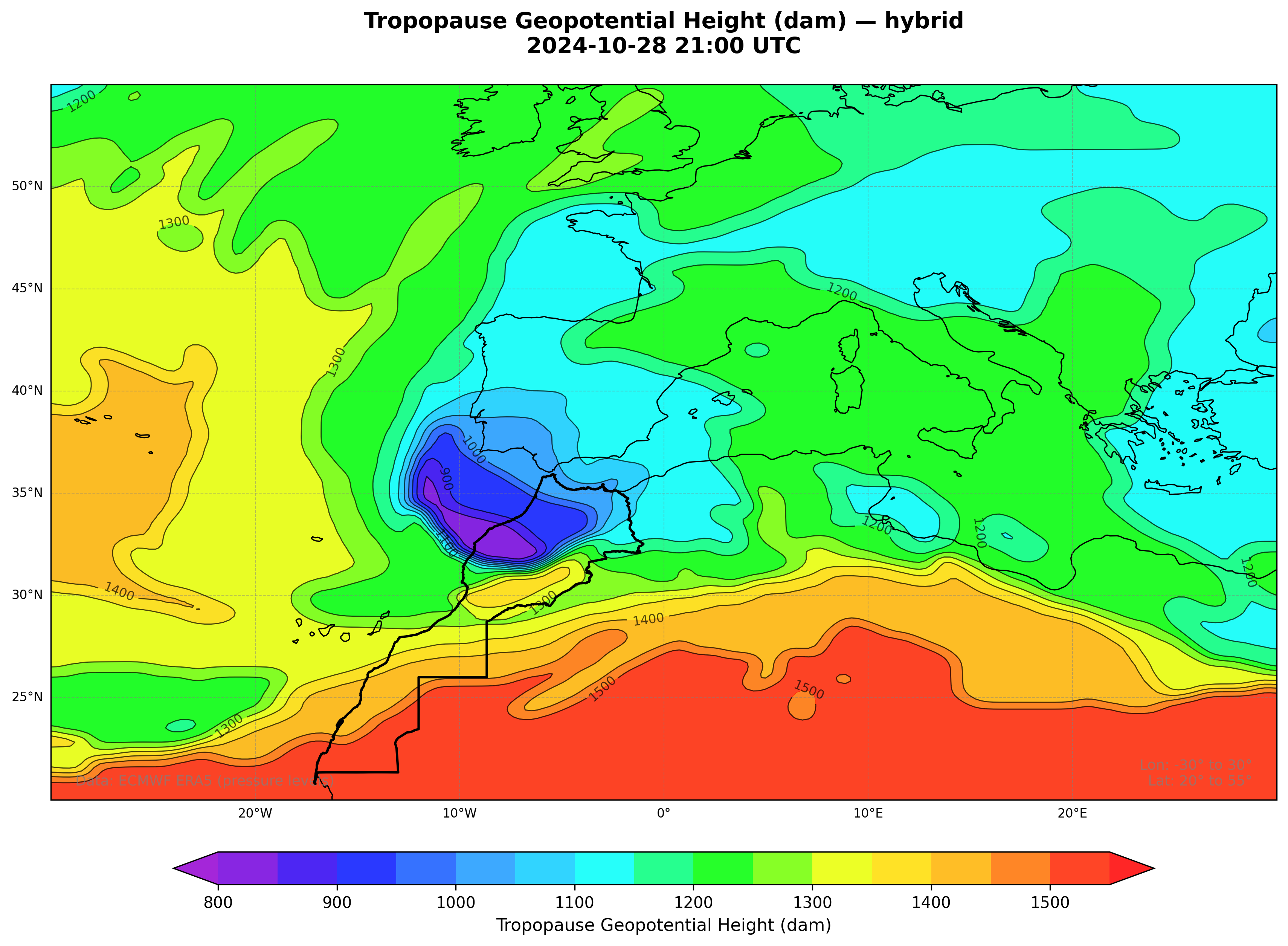}
    \label{fig:hybrid_mature}
\end{figure}

\begin{figure}[H]
    \centering
    \includegraphics[width=0.6\linewidth]{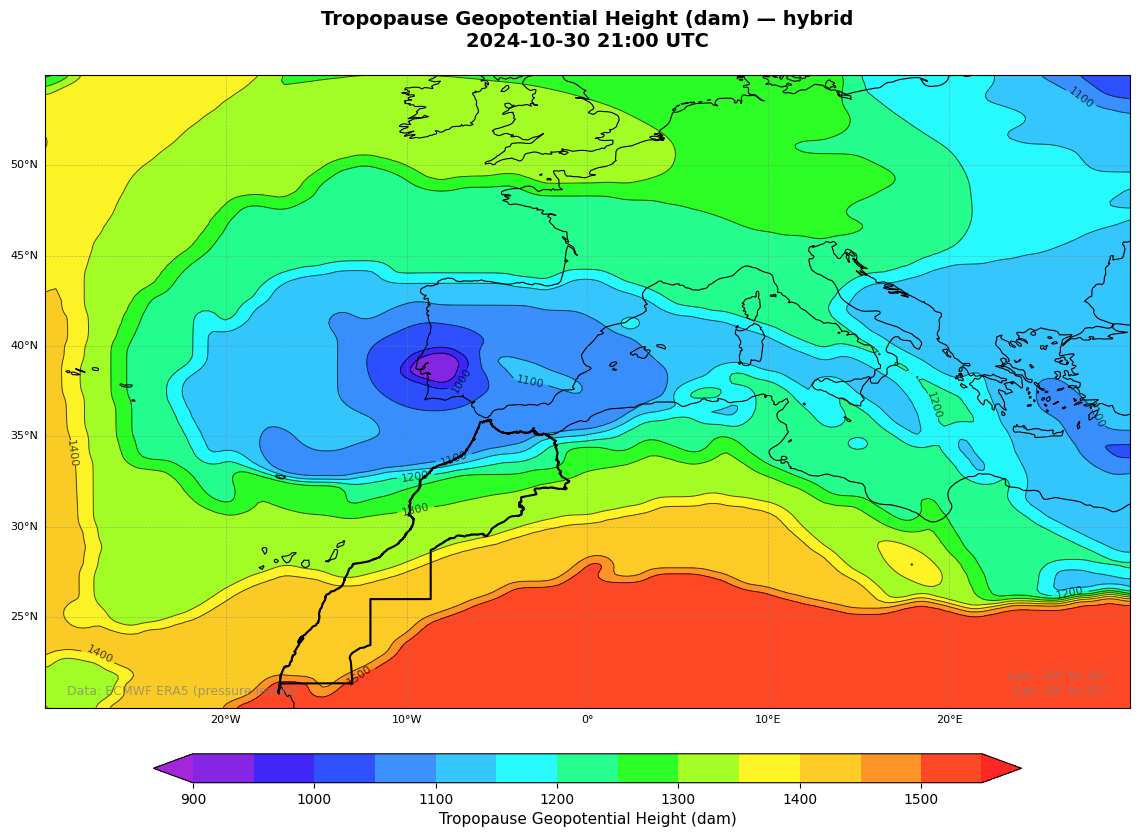}
    \label{fig:hybrid_dissipation}
\end{figure}
The geographical maps revealed key structural differences across the DANA lifecycle:
\begin{itemize}
    \item The \textbf{dynamical method} (both 1.5 PVU and 2 PVU) provided a sharp view of stratospheric intrusions throughout the event but showed less detail in thermodynamic transitions, particularly during the dissipation phase
    \item The \textbf{stability method} effectively highlighted areas of reduced tropopause height linked to atmospheric instability, showing excellent consistency during the mature phase
    \item The \textbf{hygropause method} identified strong humidity contrasts, effectively detecting moist/dry air boundaries, though performance varied with atmospheric moisture content
    \item The \textbf{hybrid method} integrated both perspectives, offering the most comprehensive representation of the tropopause structure across all stages of the DANA event
\end{itemize}

\section{Discussion}

\subsection{Strengths and Weaknesses of Each Method}
Each tropopause detection method exhibited specific strengths and limitations across the DANA lifecycle:
\begin{itemize}
    \item \textbf{Thermal (WMO) method}: Provided a consistent reference but showed discontinuities in complex atmospheric structures, particularly during the rapid development phase
    \item \textbf{Dynamical method}: Excellent for identifying stratospheric intrusions throughout the event but less sensitive to thermodynamic transitions during the dissipation phase
    \item \textbf{Stability method}: Robust in regions with complex vertical layering, showing particularly strong performance during the mature phase but occasionally overestimated tropopause height in stable conditions
    \item \textbf{Hygropause method}: Effective at detecting moisture gradients during genesis and mature phases but performed poorly in dry atmospheric conditions during dissipation
    \item \textbf{Hybrid method}: Combined the strengths of multiple approaches, demonstrating the most consistent performance across all atmospheric conditions and event phases
\end{itemize}

\subsection{Hybrid Method Performance and Implications}
The hybrid method's superior performance across all DANA lifecycle stages has important implications for atmospheric research and operational forecasting. By combining stability and humidity criteria, this approach:
\begin{itemize}
    \item Reduces false detections in complex atmospheric structures during rapid development phases
    \item Provides a more physically consistent representation of the tropopause throughout the entire event lifecycle
    \item Improves detection of stratospheric intrusions associated with cut-off lows across all development stages
    \item Enhances the ability to identify regions conducive to extreme weather development from genesis to dissipation
\end{itemize}

\subsection{Forecasting Relevance for Southern Morocco \& North Africa}
The improved tropopause detection methodology has particular relevance for forecasting in Southern Morocco and North Africa, where:
\begin{itemize}
    \item Extreme precipitation events are often triggered by upper-level dynamics associated with DANA systems
    \item In-situ meteorological observations are sparse, making satellite-based validation crucial
    \item Flash floods represent a significant natural hazard, requiring accurate early warning systems
\end{itemize}

The hybrid method's ability to accurately identify tropopause features associated with DANA events across their entire lifecycle provides valuable information for early warning systems and risk management strategies in these vulnerable regions.

The methodological framework established in this study can be operationalized in regional forecasting models, enhancing the prediction of extreme weather events and contributing to improved preparedness and resilience in North African countries.

\section{Introduction to Water Vapor 6.2$\mu$m Imagery}

The Water Vapor 6.2$\mu$m channel is particularly valuable for analyzing upper-level atmospheric moisture and dynamics. This channel is sensitive to water vapor in the upper troposphere and lower stratosphere, making it an excellent tool for identifying:
\begin{itemize}
    \item Dry intrusions (stratospheric air)
    \item Moisture patterns associated with weather systems
    \item Tropopause folds and undulations
    \item Cut-off lows and upper-level disturbances
\end{itemize}

The satellite imagery provides a unique opportunity to validate the tropopause detection methods against direct observational data.  

\begin{figure}[H]
    \centering
    \includegraphics[width=0.6\linewidth]{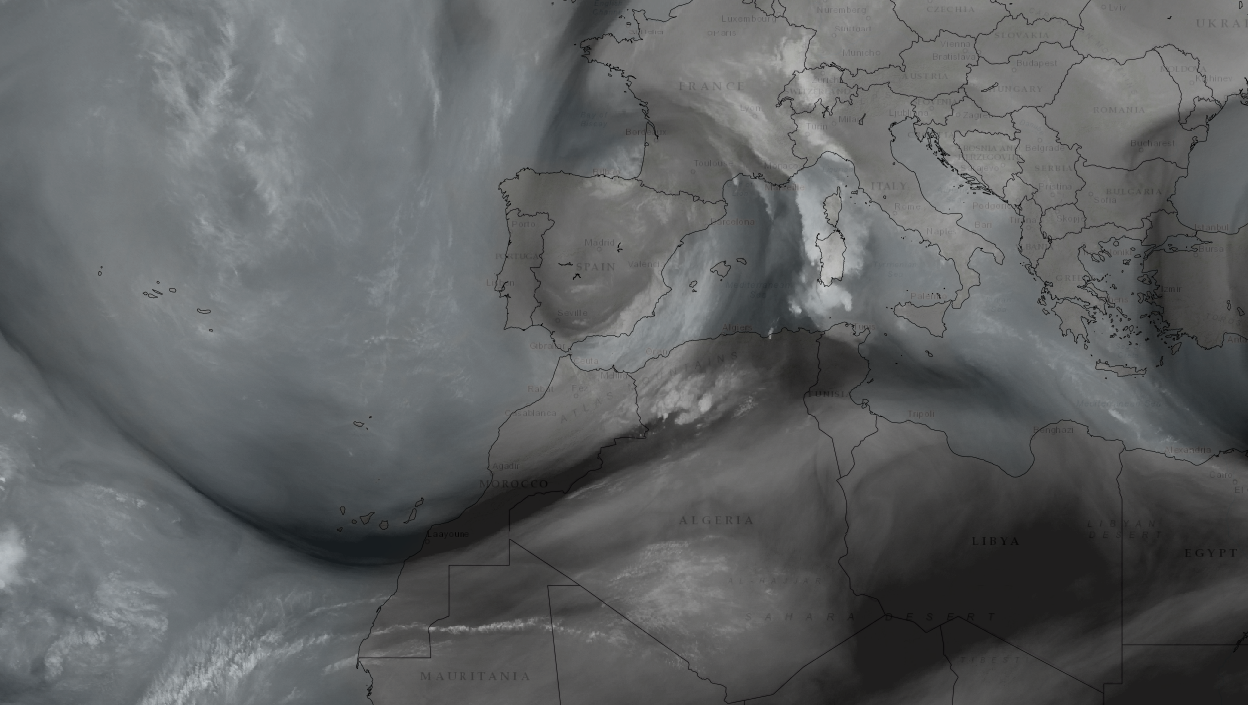}
    \caption{Water Vapor 6.2$\mu$m satellite image on 27 Oct 2024, 00:00 UTC.}
    \label{fig:wv_example_270000}
\end{figure}

\section{Automated Comparison Methodology}

A custom Python-based comparison tool was developed to systematically evaluate the performance of each tropopause detection method against water vapor satellite imagery. The comparison methodology included:

\subsection{Image Processing Pipeline}
\begin{enumerate}
    \item \textbf{Filename parsing}: Automatic extraction of timestamps from both water vapor and tropopause map filenames
    \item \textbf{Image preprocessing}: Conversion to grayscale, normalization, and resizing to ensure consistent dimensions
    \item \textbf{Similarity metrics calculation}: Computation of SSIM (Structural Similarity Index), MSE (Mean Squared Error), and NCC (Normalized Cross-Correlation)
    \item \textbf{Visual comparison}: Generation of composite figures showing original images, method outputs, and difference maps
\end{enumerate}

\subsection{Comparison Metrics}

Three quantitative metrics were employed to objectively assess the similarity between tropopause detection results and water vapor imagery:

\subsubsection{Structural Similarity Index (SSIM)}
The SSIM measures the perceptual similarity between two images, considering three key components: \textbf{luminance}, \textbf{contrast}, and \textbf{structure} \cite{ssim_paper}.

\begin{equation}
SSIM(x, y) = \frac{(2\mu_x\mu_y + C_1)(2\sigma_{xy} + C_2)}{(\mu_x^2 + \mu_y^2 + C_1)(\sigma_x^2 + \sigma_y^2 + C_2)}
\end{equation}

\begin{itemize}
    \item \textbf{$\mu_x$, $\mu_y$}: Mean intensity values of images $x$ and $y$
    \item \textbf{$\sigma_x$, $\sigma_y$}: Standard deviations of images $x$ and $y$
    \item \textbf{$\sigma_{xy}$}: Covariance between images $x$ and $y$
    \item \textbf{$C_1$, $C_2$}: Stabilization constants to avoid division by zero
\end{itemize}

\textbf{Interpretation}: 
\begin{itemize}
    \item \textbf{SSIM = 1}: Perfect structural similarity
    \item \textbf{SSIM = 0}: No structural similarity
    \item \textbf{SSIM $<$ 0}: Inverse structural relationship
    \item Values closer to 1 indicate better agreement with reference imagery
\end{itemize}

\subsubsection{Mean Squared Error (MSE)}
MSE quantifies the average squared intensity differences between corresponding pixels in two images, providing a measure of overall error magnitude \cite{mse_textbook,signal_processing_book}.

\begin{equation}
MSE = \frac{1}{MN}\sum_{i=0}^{M-1}\sum_{j=0}^{N-1}[I(i,j) - K(i,j)]^2
\end{equation}

\begin{itemize}
    \item \textbf{$M$, $N$}: Image dimensions (width and height in pixels)
    \item \textbf{$I(i,j)$}: Pixel intensity at position $(i,j)$ in reference image
    \item \textbf{$K(i,j)$}: Pixel intensity at position $(i,j)$ in compared image
    \item \textbf{$\sum$}: Summation over all pixel positions
\end{itemize}

\textbf{Interpretation}:
\begin{itemize}
    \item \textbf{MSE = 0}: Perfect pixel-wise match
    \item \textbf{MSE $>$ 0}: Increasing average squared error
    \item Lower values indicate better agreement with reference
    \item Particularly sensitive to large individual errors due to squaring
\end{itemize}

\subsubsection{Normalized Cross-Correlation (NCC)}
NCC measures the linear correlation between pixel intensities of two images, normalized to account for differences in brightness and contrast \cite{ncc_paper,pattern_recognition_book}.

\begin{equation}
NCC = \frac{\sum_{i,j}(I(i,j) - \mu_I)(K(i,j) - \mu_K)}{\sqrt{\sum_{i,j}(I(i,j) - \mu_I)^2 \sum_{i,j}(K(i,j) - \mu_K)^2}}
\end{equation}

\begin{itemize}
    \item \textbf{$\mu_I$, $\mu_K$}: Mean intensity values of reference and compared images
    \item \textbf{$I(i,j) - \mu_I$}: Deviation from mean in reference image
    \item \textbf{$K(i,j) - \mu_K$}: Deviation from mean in compared image
    \item The numerator represents covariance, denominator normalizes by standard deviations
\end{itemize}

\textbf{Interpretation}:
\begin{itemize}
    \item \textbf{NCC = 1}: Perfect positive linear correlation
    \item \textbf{NCC = -1}: Perfect negative linear correlation (inverse relationship)
    \item \textbf{NCC = 0}: No linear correlation
    \item Values approaching 1 indicate strong agreement in patterns and structures
\end{itemize}

\subsection{Comparison Time Period}
The analysis covers the DANA event from 27--29 October 2024, with 3-hourly comparisons at the following times:
\begin{itemize}
    \item 27 October: 00:00, 06:00, 12:00, 18:00 UTC
    \item 28 October: 00:00, 06:00, 12:00, 18:00 UTC
    \item 29 October: 00:00, 06:00, 12:00, 18:00 UTC
    \item 30 October: 00:00 UTC
\end{itemize}
\section{Dynamical Tropopause Definition}

For this study, the dynamical tropopause was defined using the \textbf{1.5 PVU} isosurface, which provides enhanced sensitivity for detecting stratospheric intrusions during cut-off low situations. This threshold was selected based on its better performance in capturing fine-scale features associated with DANA events compared to the traditional 2 PVU definition.

\section{Comparative Analysis Results}

\subsection{Visual Comparison Outputs}

This section presents visual comparisons between water vapor (WV) images and tropopause detection methods across three distinct meteorological phases: genesis, mature phase, and dissipation phase. For each phase, representative timesteps are analyzed to evaluate the performance of different detection methods.

\subsubsection{Genesis Phase}

\begin{figure}[H]
    \centering
    \includegraphics[width=0.8\textwidth]{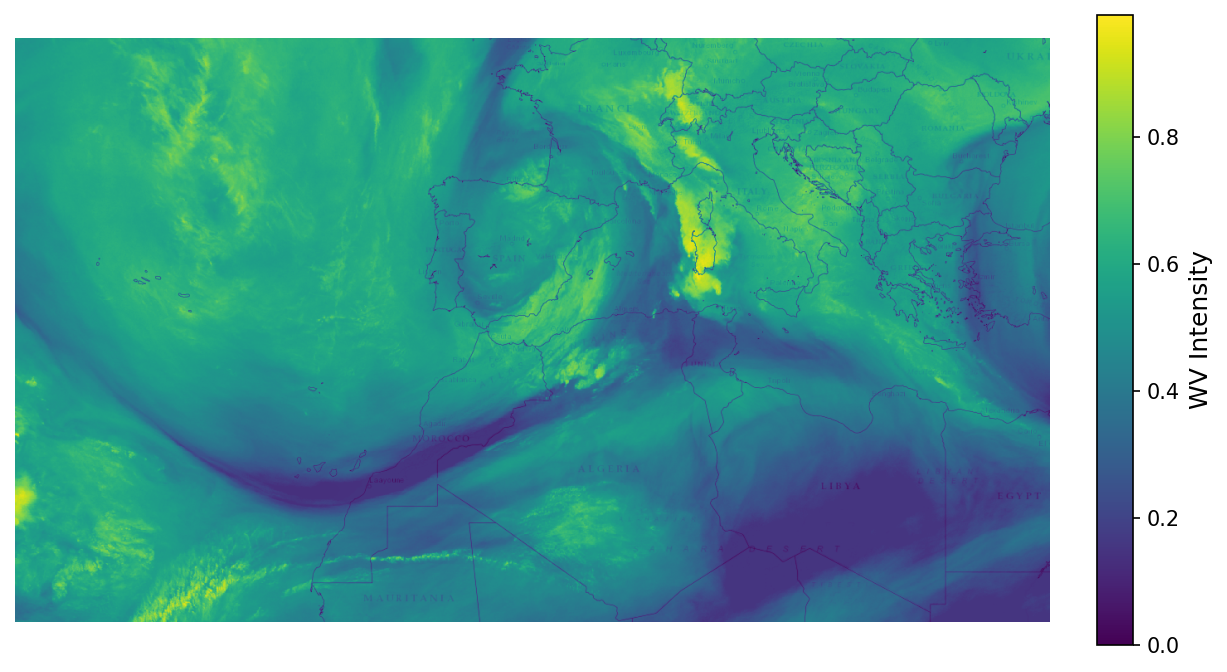}
    \caption{Water Vapor Image during genesis phase (Timestamp: 20241027\_0000). Shows the initial development of atmospheric features.}
    \label{fig:genesis_wv}
\end{figure}

\begin{figure}[H]
    \centering
    \includegraphics[width=1\textwidth]{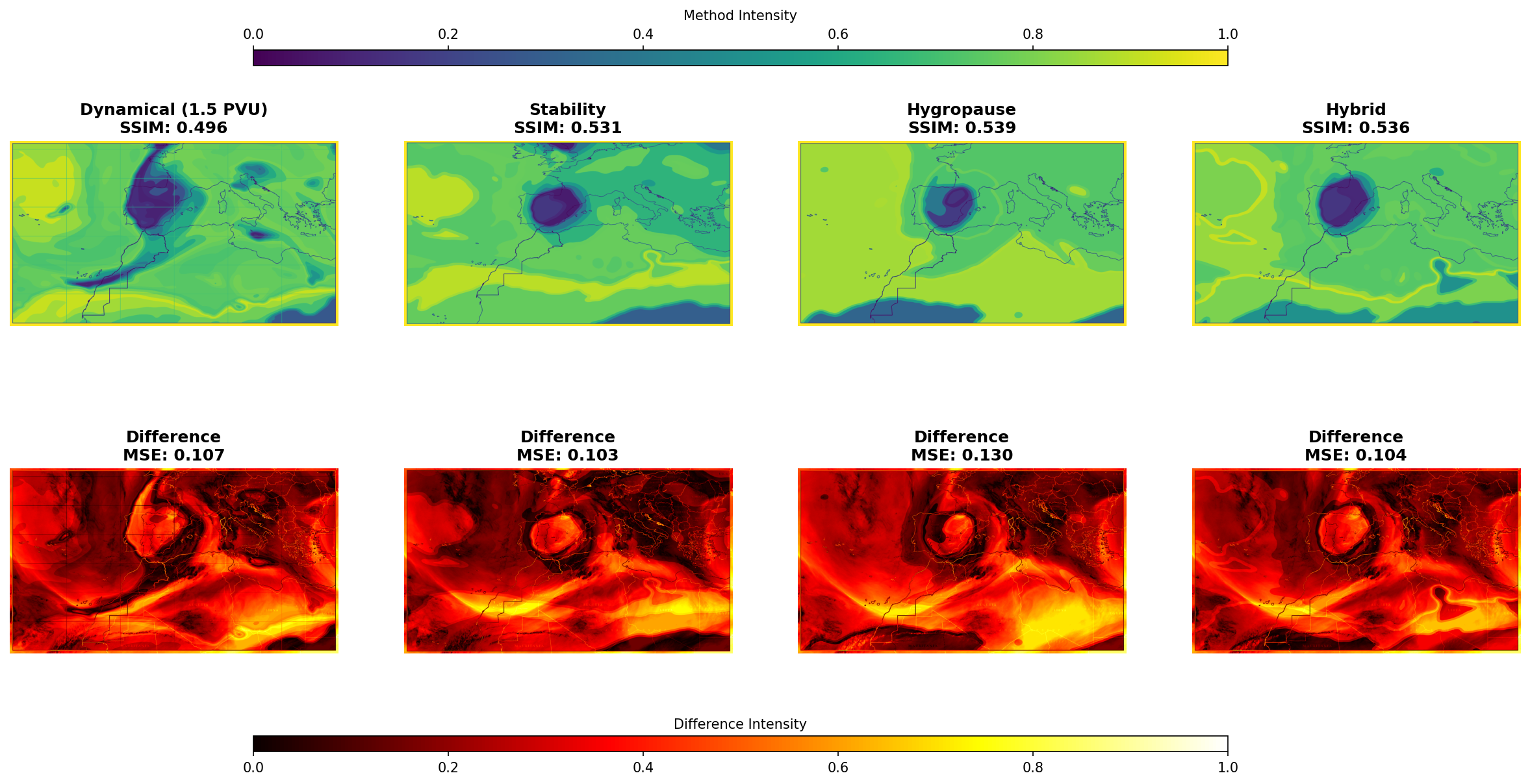}
    \caption{Method comparison during genesis phase (Timestamp: 20241027\_0000). Top row shows tropopause detection methods, bottom row shows difference maps with WV image.}
    \label{fig:genesis_comparison}
\end{figure}
\vspace{0.5cm}
Figure \ref{fig:genesis_wv} shows the water vapor image during the genesis phase, displaying the initial development of atmospheric features. Figure \ref{fig:genesis_comparison} demonstrates the performance of various detection methods and their differences from the WV reference during this early development stage.

\subsubsection{Mature Phase}

\begin{figure}[H]
    \centering
    \includegraphics[width=0.8\textwidth]{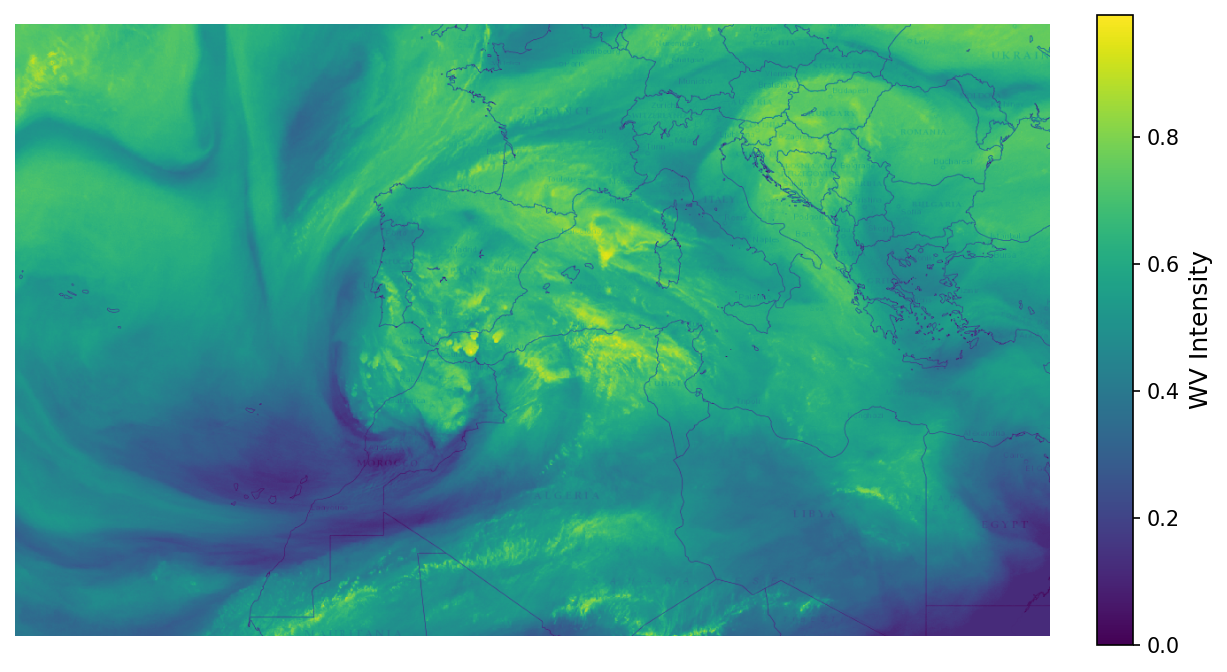}
    \caption{Water Vapor Image during mature phase (Timestamp: 20241028\_1200). The fully developed atmospheric system shows distinct tropopause features.}
    \label{fig:mature_wv}
\end{figure}

\begin{figure}[H]
    \centering
    \includegraphics[width=1\textwidth]{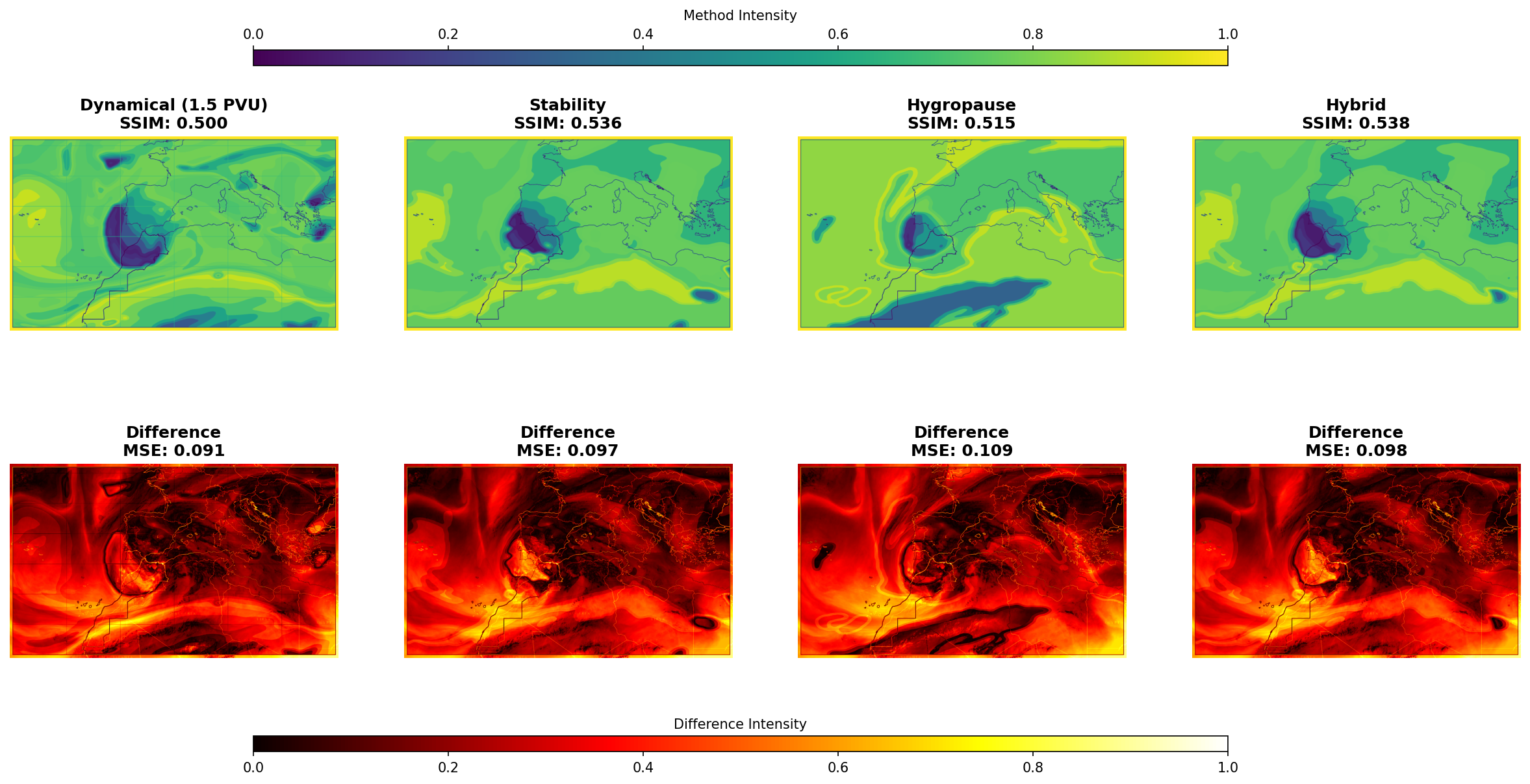}
    \caption{Method comparison during mature phase (Timestamp: 20241028\_1200). Performance evaluation under well-developed atmospheric conditions.}
    \label{fig:mature_comparison}
\end{figure}
\vspace{0.5cm}
During the mature phase, the atmospheric system reaches its peak development. Figure \ref{fig:mature_wv} exhibits well-defined structures in the water vapor image, providing a robust benchmark for evaluating method performance in Figure \ref{fig:mature_comparison}.

\subsubsection{Dissipation Phase}

\begin{figure}[H]
    \centering
    \includegraphics[width=0.8\textwidth]{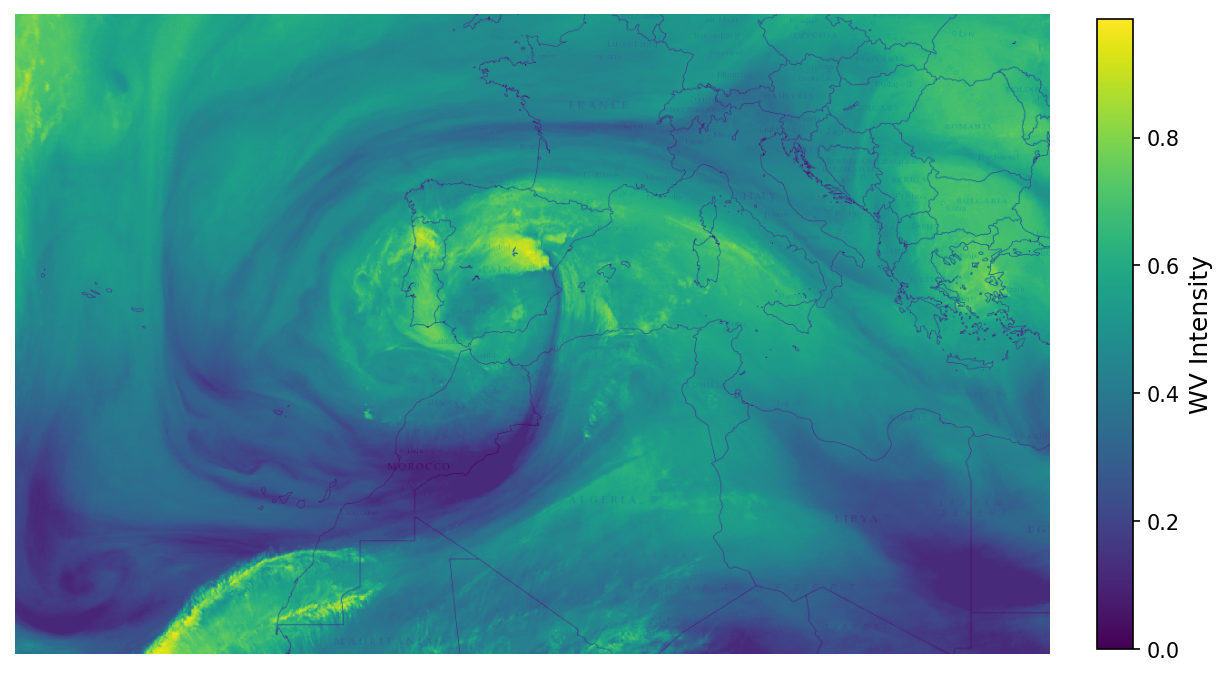}
    \caption{Water Vapor Image during dissipation phase (Timestamp: 20241029\_1800). Shows the decaying atmospheric system with diminishing structural definition.}
    \label{fig:dissipation_wv}
\end{figure}

\begin{figure}[H]
    \centering
    \includegraphics[width=1\textwidth]{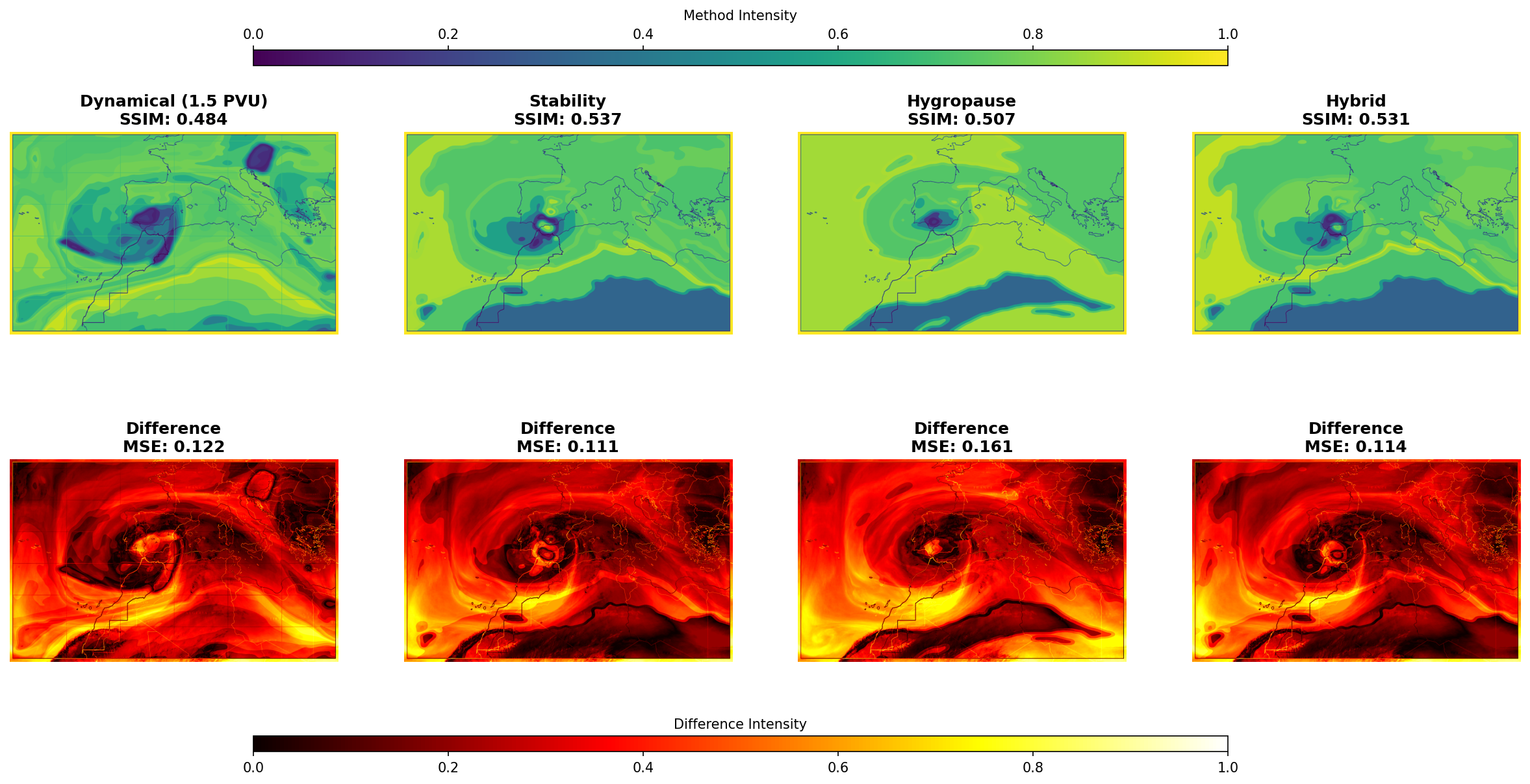}
    \caption{Method comparison during dissipation phase (Timestamp: 20241029\_1800). Tests the robustness of detection methods under challenging decaying conditions.}
    \label{fig:dissipation_comparison}
\end{figure}
\vspace{0.5cm}
Figure \ref{fig:dissipation_wv} illustrates the water vapor image during the dissipation phase, where the atmospheric system undergoes decay with diminishing structural definition. Figure \ref{fig:dissipation_comparison} tests the robustness of detection methods under these challenging conditions.

\subsection{Quantitative Performance Metrics}

Table \ref{tab:performance_metrics} summarizes the quantitative performance metrics (SSIM, MSE, NCC) across the three key atmospheric phases, providing a comprehensive assessment of each method's effectiveness.

\begin{table}[H]
\centering

\begin{tabular}{lcccc}
\toprule
\textbf{Method} & \textbf{Phase} & \textbf{SSIM} & \textbf{MSE} & \textbf{NCC} \\
\midrule
\multirow{3}{*}{Dynamical (1.5 PVU)} 
 & Genesis & 0.4965 & 0.1071 & 0.0853 \\
 & Mature & 0.4998 & 0.0909 & -0.0071 \\
 & Dissipation & 0.4839 & 0.1222 & -0.0429 \\
\cmidrule(lr){1-5}
\multirow{3}{*}{Stability}
 & Genesis & 0.5308 & 0.1035 & 0.0261 \\
 & Mature & 0.5364 & 0.0968 & -0.1311 \\
 & Dissipation & 0.5374 & 0.1106 & 0.0750 \\
\cmidrule(lr){1-5}
\multirow{3}{*}{Hygropause}
 & Genesis & 0.5391 & 0.1304 & -0.0440 \\
 & Mature & 0.5151 & 0.1086 & -0.0518 \\
 & Dissipation & 0.5067 & 0.1609 & -0.1535 \\
\cmidrule(lr){1-5}
\multirow{3}{*}{Hybrid}
 & Genesis & 0.5362 & 0.1036 & 0.1008 \\
 & Mature & 0.5376 & 0.0977 & -0.1393 \\
 & Dissipation & 0.5313 & 0.1145 & 0.1105 \\
\bottomrule
\end{tabular}
\caption{Performance metrics across different atmospheric phases}
\label{tab:performance_metrics}
\end{table}

\begin{table}[H]
\centering
\begin{tabular}{lccc}
\toprule
\textbf{Method} & \textbf{Mean SSIM} & \textbf{Mean MSE} & \textbf{Mean NCC} \\
\midrule
Dynamical (1.5 PVU) & 0.4900 $\pm$ 0.0114 & 0.1197 $\pm$ 0.0203 & 0.0274 $\pm$ 0.0306 \\
Stability & 0.5298 $\pm$ 0.0167 & 0.1204 $\pm$ 0.0210 & -0.0503 $\pm$ 0.1351 \\
Hygropause & 0.5166 $\pm$ 0.0145 & 0.1488 $\pm$ 0.0254 & -0.0482 $\pm$ 0.0392 \\
Hybrid & 0.5315 $\pm$ 0.0137 & 0.1181 $\pm$ 0.0155 & 0.0397 $\pm$ 0.1573 \\
\bottomrule
\end{tabular}
\caption{Overall comparison metrics between water vapor imagery and tropopause detection maps (averages over 27--30 October 2024).}
\label{tab:comparison_metrics}
\end{table}
\newpage
The quantitative results reveal a more nuanced performance pattern across methods. While the Hybrid method shows competitive performance with the highest overall mean SSIM (0.5315) and lowest mean MSE (0.1181), the Stability method actually achieves the best performance in the Dissipation phase with the highest SSIM (0.5374) and relatively low MSE (0.1106). Interestingly, the Hygropause method shows the strongest performance during the Genesis phase (SSIM: 0.5391) despite higher MSE values. 

All methods demonstrate variable performance across different atmospheric phases, with no single method consistently dominating across all metrics and phases. The mature phase generally shows more stable performance across methods, while greater variability is observed during both genesis and dissipation phases, reflecting the challenges of tropopause detection during developing and decaying atmospheric systems.

\section{Time Series and Spatial Pattern Analysis}

\begin{figure}[H]
    \centering
    \includegraphics[width=0.8\linewidth]{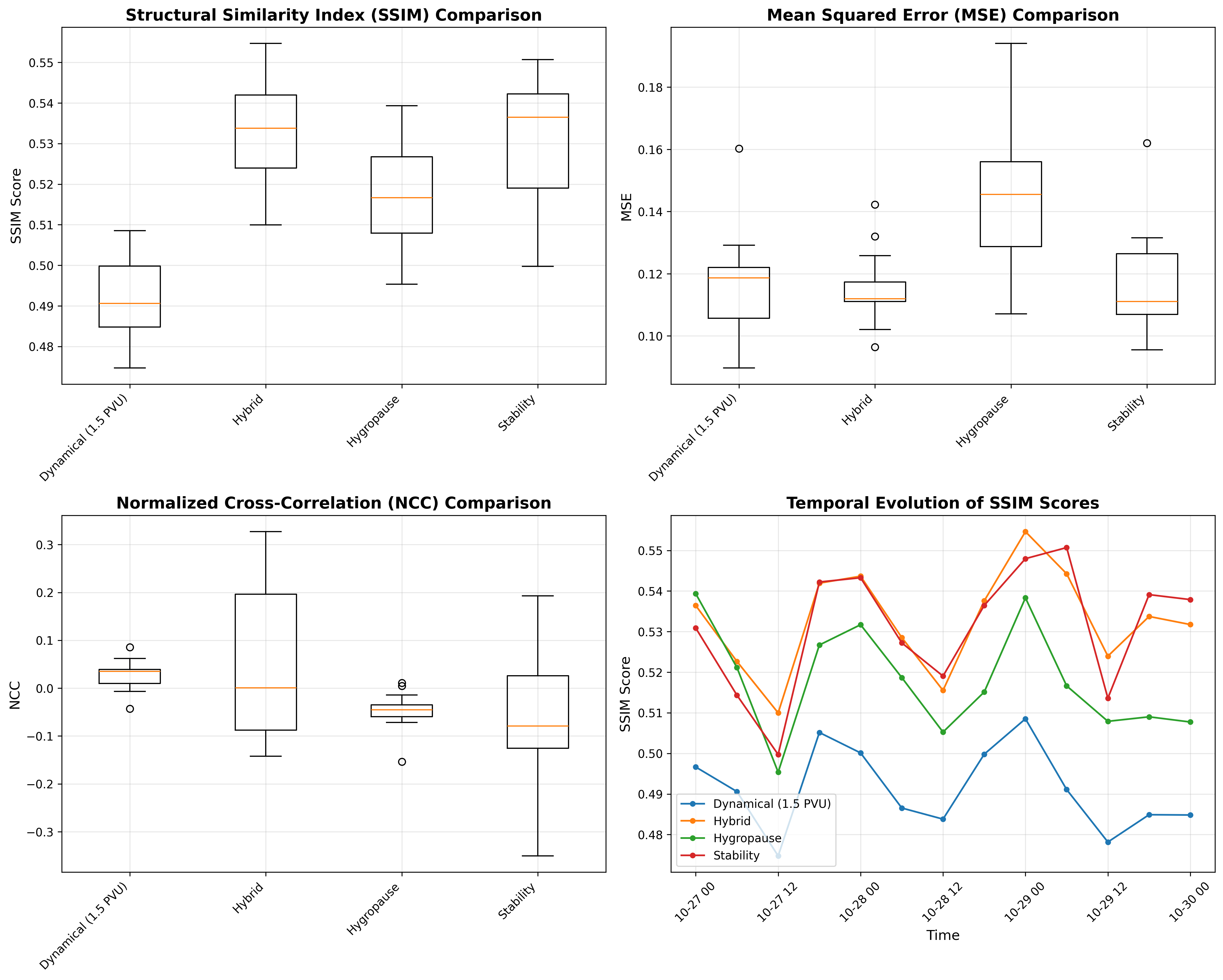} 
    \caption{Comprehensive comparison of similarity metrics}
    \label{fig:metrics_comprehensive}
\end{figure}

\vspace{2cm}
The boxplots show that the hybrid and stability methods generally achieve higher SSIM values, while the dynamical method tends to underperform.  
MSE indicates larger variability for the hygropause criterion, and NCC highlights that the hybrid approach captures stronger correlations but with broader dispersion.  
\vspace{2cm}
\begin{figure}[H]
    \centering
    \includegraphics[width=0.9\linewidth]{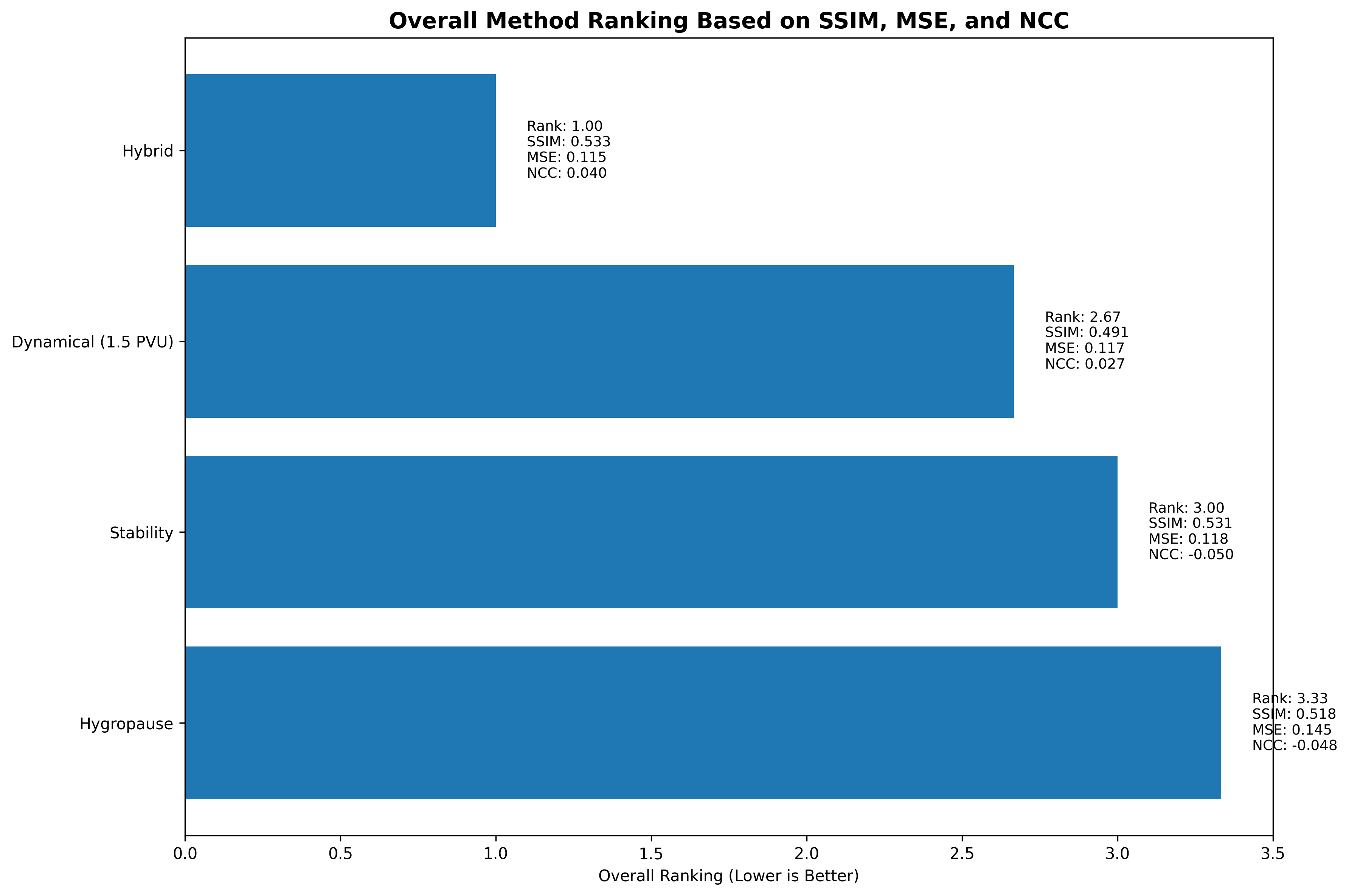}
    \caption{Overall Method Ranking Based on SSIM, MSE, and NCC}
    \label{fig:method_ranking}
\end{figure}
\v
The ranking confirms the hybrid method as the best compromise, followed by the dynamical criterion, while stability and hygropause show greater variability in performance.

\chapter{Conclusion and Perspectives}
\section{Statistical Summary}

\begin{table}[H]
    \centering
    \begin{tabular}{lccc}
        \toprule
        Method & Mean SSIM & Std. Dev. & 95\% CI \\
        \midrule
        Dynamical (1.5 PVU) & 0.4911 & 0.0099 & [0.4747--0.5085] \\
        Stability           & 0.5310 & 0.0147 & [0.4997--0.5507] \\
        Hygropause          & 0.5179 & 0.0128 & [0.4954--0.5393] \\
        Hybrid              & 0.5327 & 0.0120 & [0.5100--0.5546] \\
        \bottomrule
    \end{tabular}
    \caption{Statistical summary of comparative metrics (27--30 October 2024). 95\% confidence intervals are shown.}
    \label{tab:stats_summary}
\end{table}

\section{Conclusion}

This study investigated multiple approaches for tropopause detection during the 
DANA event of 27--30 October 2024, comparing them systematically against 
satellite observations from the Water Vapor 6.2$\mu$m channel.  

The analysis confirmed that:
\begin{itemize}
    \item The \textbf{hybrid method} provided the best overall agreement with water vapor imagery, showing consistently high SSIM scores and relatively low error values.
    \item The \textbf{stability criterion} performed well in several instances, especially during dynamically active phases, but showed greater variability.
    \item The \textbf{dynamical (1.5 PVU) tropopause} captured the main synoptic structures but tended to underperform compared to hybrid and stability methods in terms of similarity metrics.
    \item The \textbf{hygropause criterion} displayed the largest variability and lower correlations, reflecting its sensitivity to missing or less reliable humidity information.
\end{itemize}

These results highlight the added value of combining multiple physical definitions of the tropopause. The hybrid approach, in particular, emerges as a robust tool for detecting fine-scale dynamical structures and improving consistency with satellite water vapor imagery.

\bigskip
\begin{quote}
    \textit{``The meteorologist is like an artist-painter: they must know how to arrange the colours and handle the brush to realistically depict a subject.''}
    \vspace{0.5em}\\
    \hfill\emph{Pierre Gauthier (UQAM), renowned expert in data assimilation}
\end{quote}
This statement resonates deeply with the work presented here, as it reflects the intricate balance of combining various methods and tools to provide a detailed, accurate portrait of the tropopause in the context of the DANA event.

\section{Perspectives}

Several perspectives can be drawn from this work:
\begin{itemize}
    \item \textbf{Extension to larger datasets}: Applying the methodology to longer time series and diverse meteorological events would provide a broader assessment of method robustness across different seasons and regions.
    \item \textbf{Refinement of the hybrid approach}: The weighting strategy between dynamical, stability, and hygropause criteria could be optimized to improve adaptability to specific atmospheric conditions.
    \item \textbf{Integration of additional observational datasets}: Combining satellite water vapor imagery with ozone profiles, reanalysis humidity fields, or aircraft observations could enhance validation and reduce uncertainty.
    \item \textbf{Preoperational implementation}: Incorporating the automated comparison pipeline into forecasting workflows could provide real-time diagnostics of tropopause dynamics during high-impact weather events.
    \item \textbf{Machine learning applications}: Future work may explore data-driven approaches to learn optimal detection thresholds and improve generalization across different atmospheric regimes.
    \item \textbf{Application to high-resolution models}: Since the present study relies on ERA5 reanalysis data, applying the same methodology to convection-permitting models such as AROME would be highly valuable. This would allow testing the detection criteria on finer spatial and temporal scales, and assessing their suitability for operational forecasting in complex terrain regions.
\end{itemize}

Overall, the methodology developed here demonstrates the feasibility of quantitative, observation-based evaluation of tropopause detection methods, offering a strong foundation for both research applications and operational forecasting improvements.

\chapter{Annexe}

\section{Data Acquisition and Processing Scripts}

This appendix contains the complete code and scripts used for downloading and processing ERA5 data for this study.

\subsection{Linux Environment Setup}

The following commands were used to set up the Python environment on Ubuntu Linux:

\begin{verbatim}
# Update system packages
sudo apt update
sudo apt install python3 python3-pip python3-venv

# Create and activate virtual environment
python3 -m venv ~/era5_env
source ~/era5_env/bin/activate

# Install required Python packages
pip install cdsapi xarray netCDF4 numpy matplotlib pandas
\end{verbatim}

\subsection{ECMWF API Configuration}

The ECMWF API requires authentication credentials stored in \texttt{\textasciitilde/.cdsapirc}:

\begin{verbatim}
url: https://cds.climate.copernicus.eu/api/v2
key: 123456:abcdefghijklmn-opqrstuvwxyz0123456789
\end{verbatim}

\subsection{Data Download Script}

The main Python script for downloading ERA5 data:

\begin{verbatim}
import cdsapi
import datetime
import os

def download_era5_data():
    c = cdsapi.Client()
    
    # Define study periods
    date_ranges = [
        '2010-01-01/to/2010-01-31',  # Monthly validation
        '2024-09-06/to/2024-09-07',  # Morocco events
        '2024-09-20/to/2024-09-21',
        '2024-10-27/to/2024-10-30',  # DANA event
        '2025-01-27/to/2025-01-28'   # Morocco event
    ]
    
    for i, date_range in enumerate(date_ranges):
        request = {
            'product_type': 'reanalysis',
            'format': 'netcdf',
            'variable': [
                'temperature', 'geopotential', 'specific_humidity',
                'potential_vorticity', 'relative_humidity'
            ],
            'pressure_level': [
                '1', '2', '3', '5', '7', '10', '20', '30', '50',
                '70', '100', '125', '150', '175', '200', '225',
                '250', '300', '350', '400', '450', '500', '550',
                '600', '650', '700', '750', '775', '800', '825',
                '850', '875', '900', '925', '950', '975', '1000'
            ],
            'date': date_range,
            'time': ['00:00', '06:00', '12:00', '18:00'],
            'area': [40, -20, 20, 10],  # N, W, S, E
        }
        
        output_file = f'era5_data_{i}.nc'
        c.retrieve('reanalysis-era5-pressure-levels', request, output_file)
        print(f'Downloaded {output_file}')

if __name__ == '__main__':
    download_era5_data()
\end{verbatim}

\subsection{Data Processing Script}

Python script for processing and organizing the downloaded data:

\begin{verbatim}
import xarray as xr
import pandas as pd
import os
from datetime import datetime, timedelta

def process_era5_data():
    # Create organized directory structure
    os.makedirs('data/era5/2010_01', exist_ok=True)
    os.makedirs('data/era5/2024_09', exist_ok=True)
    os.makedirs('data/era5/2024_10', exist_ok=True)
    os.makedirs('data/era5/2025_01', exist_ok=True)
    
    # Process each downloaded file
    for i in range(5):
        input_file = f'era5_data_{i}.nc'
        
        try:
            ds = xr.open_dataset(input_file)
            
            # Extract dates and organize by event
            dates = pd.to_datetime(ds.time.values)
            
            for date in dates:
                date_str = date.strftime('%Y%m%d')
                daily_data = ds.sel(time=date)
                
                # Determine output directory based on date
                if date.year == 2010:
                    output_dir = 'data/era5/2010_01'
                elif date.month == 9 and date.year == 2024:
                    output_dir = 'data/era5/2024_09'
                elif date.month == 10 and date.year == 2024:
                    output_dir = 'data/era5/2024_10'
                elif date.year == 2025:
                    output_dir = 'data/era5/2025_01'
                
                # Save daily file
                output_file = f'{output_dir}/era5_{date_str}.nc'
                daily_data.to_netcdf(output_file)
                
            ds.close()
            print(f'Processed {input_file}')
            
        except Exception as e:
            print(f'Error processing {input_file}: {e}')

def quality_control():
    """Perform basic quality control checks"""
    print("Performing quality control checks...")
    
    # Check for missing values
    for root, dirs, files in os.walk('data/era5'):
        for file in files:
            if file.endswith('.nc'):
                filepath = os.path.join(root, file)
                try:
                    ds = xr.open_dataset(filepath)
                    
                    # Check for NaN values
                    for var in ds.data_vars:
                        nan_count = ds[var].isnull().sum().values
                        if nan_count > 0:
                            print(f"Warning: {var} in {file} has {nan_count} NaN values")
                    
                    ds.close()
                except:
                    print(f"Could not open {filepath}")

if __name__ == '__main__':
    process_era5_data()
    quality_control()
\end{verbatim}

\subsection{Bash Automation Script}

Shell script to automate the entire process:

\begin{verbatim}
#!/bin/bash

# ERA5 Data Download and Processing Automation Script
# Author: Mohammed El Abdioui
# Date: $(date +%Y-%m-%d)

echo "Starting ERA5 data processing pipeline..."
echo "==========================================="

# Activate Python environment
source ~/era5_env/bin/activate

echo "1. Downloading ERA5 data..."
python download_era5.py

echo "2. Processing and organizing data..."
python process_era5.py

echo "3. Performing quality control..."
python quality_control.py

echo "4. Generating data inventory..."
find data/era5 -name "*.nc" | wc -l > data/file_count.txt
du -sh data/era5/ > data/directory_size.txt

echo "Processing complete!"
echo "Summary:"
cat data/file_count.txt
cat data/directory_size.txt
\end{verbatim}

\subsection{Full Project Code}

All the scripts, code, and the complete report for this project are available on my GitHub repository:

\begin{center}
\url{https://github.com/simoghost99/Tropopause/tree/main}
\end{center}


\begin{thebibliography}{9}

\bibitem{ecmwf_stratosphere}
ECMWF. (2024). \textit{Reintroducing analysis humidity in the stratosphere}.
Retrieved from \url{https://www.ecmwf.int/en/newsletter/183/earth-system-science/reintroducing-analysis-humidity-stratosphere}

\bibitem{ifs_documentation}
ECMWF. (2024). \textit{IFS Documentation - Cy49r1, Part IV: Physical Processes}.
Operational implementation 12 November 2024.
Retrieved from \url{https://www.ecmwf.int/sites/default/files/elibrary/112024/81626-ifs-documentation-cy49r1-part-iv-physical-processes.pdf}

\bibitem{limb_correction}
Elmer, N. J., Berndt, E., Jedlovec, G., \& Fuell, K. (2019). 
\textit{Limb Correction of Geostationary Infrared Imagery in Clear and Cloudy Regions to Improve Interpretation of RGB Composites for Real-Time Applications}.
Journal of Atmospheric and Oceanic Technology, 36(8), 1675-1690.
DOI: \url{https://doi.org/10.1175/JTECH-D-18-0206.1}

\bibitem{potential_vorticity}
Hoskins, B. J., McIntyre, M. E., \& Robertson, A. W. (1985). 
\textit{On the use and significance of isentropic potential vorticity maps}.
Quarterly Journal of the Royal Meteorological Society, 111(470), 877-946.
DOI: \url{https://doi.org/10.1002/qj.49711147002}

\bibitem{era5_documentation}
ECMWF. (2024). \textit{ERA5: Fifth generation of ECMWF atmospheric reanalyses of the global climate}.
Retrieved from \url{https://www.ecmwf.int/en/forecasts/datasets/reanalysis-datasets/era5}

\bibitem{ecmwf_website}
European Centre for Medium-Range Weather Forecasts (ECMWF). (2024).
\textit{Official website}. Retrieved from \url{https://www.ecmwf.int}

\bibitem{santer2004}
Santer, B. D., Wehner, M. F., Wigley, T. M. L., Sausen, R., Meehl, G. A., 
Taylor, K. E., \ldots \& Branson, M. (2004). 
\textit{Contributions of anthropogenic and natural forcing to recent tropopause height changes}.
Science, 301(5632), 479-483.
DOI: \url{https://doi.org/10.1126/science.1084123}

\bibitem{wmo_tropopause}
World Meteorological Organization (WMO). (1957). 
\textit{Definition of the tropopause}.
WMO Bulletin, 6(4), 136-137.

\bibitem{dynamic_tropopause}
Hoinka, K. P. (1998). 
\textit{Statistics of the global tropopause pressure}.
Monthly Weather Review, 126(12), 3303-3325.
DOI: \url{https://doi.org/10.1175/1520-0493(1998)126<3303:SOTGTP>2.0.CO;2}

\bibitem{hybrid_method}
ECMWF. (2023). \textit{Hybrid tropopause detection methods in the IFS}.
ECMWF Technical Memorandum No. 987.
\bibitem{ssim_paper}
Wang, Z., Bovik, A. C., Sheikh, H. R., \& Simoncelli, E. P. (2004). 
\textit{Image quality assessment: From error visibility to structural similarity}. 
IEEE Transactions on Image Processing, 13(4), 600-612.
DOI: \url{https://doi.org/10.1109/TIP.2003.819861}

\bibitem{mse_textbook}
Gonzalez, R. C., \& Woods, R. E. (2018). 
\textit{Digital Image Processing} (4th ed.). 
Pearson Education.
Chapter 5: Image Restoration and Reconstruction.

\bibitem{ncc_paper}
Lewis, J. P. (1995). 
\textit{Fast normalized cross-correlation}. 
Vision Interface, 10(1), 120-123.

\bibitem{image_metrics_review}
Hore, A., \& Ziou, D. (2010). 
\textit{Image quality metrics: PSNR vs. SSIM}. 
In 2010 20th International Conference on Pattern Recognition (pp. 2366-2369). IEEE.
DOI: \url{https://doi.org/10.1109/ICPR.2010.579}

\bibitem{pattern_recognition_book}
Theodoridis, S., \& Koutroumbas, K. (2008). 
\textit{Pattern Recognition} (4th ed.). 
Academic Press.
Chapter 7: Feature Generation II.

\bibitem{computer_vision_book}
Szeliski, R. (2010). 
\textit{Computer Vision: Algorithms and Applications}. 
Springer Science \& Business Media.
Chapter 8: Dense motion estimation.

\bibitem{signal_processing_book}
Oppenheim, A. V., \& Schafer, R. W. (2009). 
\textit{Discrete-Time Signal Processing} (3rd ed.). 
Pearson Education.
Chapter 7: Filter Design Techniques.
\end{thebibliography}
\end{document}